\documentclass[a4paper,fleqn,11pt]{article}

\usepackage[numbers]{natbib}

\usepackage{amsmath,amssymb,amsfonts}
\usepackage{algorithmic}
\usepackage{graphicx}
\usepackage{textcomp}
\usepackage{bmpsize}
\usepackage{xcolor}
\usepackage{lipsum}

\usepackage{amsmath,amsfonts}
\usepackage{algorithmic}
\usepackage{array}
\usepackage[caption=false,font=normalsize,labelfont=sf,textfont=sf]{subfig}
\usepackage{textcomp}
\usepackage{stfloats}
\usepackage{url}
\usepackage{verbatim}
\usepackage{graphicx}

\usepackage[table]{xcolor}

\usepackage[utf8]{inputenc}

\usepackage{booktabs}
\usepackage{rotating}
\usepackage{pdflscape}
\usepackage{afterpage}
\usepackage{adjustbox}

\makeatletter
\newcommand\notsotiny{\@setfontsize\notsotiny\@vipt\@viipt}
\makeatother

\def\BibTeX{{\rm B\kern-.05em{\sc i\kern-.025em b}\kern-.08em
    T\kern-.1667em\lower.7ex\hbox{E}\kern-.125emX}}

\usepackage{gensymb}
\usepackage{amssymb}
\usepackage[font=singlespacing]{caption}

\usepackage{tikz}
\usepackage{amsmath}

\newcommand*\circled[1]{\tikz[baseline=(char.base)]{
    \node[shape=circle,draw,fill=black,text=white,inner sep=1pt,outer sep=0pt,minimum size=1.2em] (char) {#1};}}

\usepackage{filecontents}

\usepackage{comment}
\usepackage{algorithmic}
\usepackage{graphicx}
\usepackage{textcomp}
\usepackage{xcolor}
\usepackage{float}
\usepackage{pgfplots}
\pgfplotsset{compat=1.16}
\usepackage{diagbox}

\usepackage{fancyhdr}

\usepackage [ T1 ] { fontenc }
\usepackage [ linguistics ] {forest}

\usepackage[caption=false]{subfig}

\forestset{%
  angle below/.style={
    tikz+={%
      \draw ($(!1.child anchor)!.35!(.parent anchor)$) [bend right=3.33] to ($(.parent anchor)!.65!(!l.child anchor)$);
    },
  },
}

\forestset{%
  angle below last two/.style={
    tikz+={%
      \draw ($(!2.child anchor)!.35!(.parent anchor)$) [bend right=3.33] to ($(.parent anchor)!.65!(!l.child anchor)$);
    },
  },
}

\forestset{%
  angle below skip first two/.style={
    tikz+={%
      \draw ($(!3.child anchor)!.35!(.parent anchor)$) [bend right=3.33] to ($(.parent anchor)!.65!(!l.child anchor)$);
    },
  },
}

\newcommand{\fullcircle}{{\color{green!80!black}\tikz\fill[black] (0,0) circle (.8ex);}}
\newcommand{\emptycircle}{{\color{red!80!black}\tikz\draw[black, thick] (0,0) circle (.8ex);}}
\newcommand{\halfcircle}{{\color{orange!80!black}\tikz[baseline=-0.5ex]\draw[black, thick] (0,0) circle (.8ex) arc (90:270:.8ex);}}

\usepackage{stackengine}

\usepackage{tikz}
\newcommand{\halfcirc}{%
  \tikz[baseline=-0.5ex]{%
    \draw (0,0) circle (0.42ex);
    \fill (0,0) -- (90:0.42ex) arc (90:270:0.42ex) -- cycle;
  }%
}

\begin{document}
\let\WriteBookmarks\relax
\def\floatpagepagefraction{1}
\def\textpagefraction{.001}

\title{Evasion Adversarial Attacks Remain Impractical Against ML-based Network Intrusion Detection Systems, Especially Dynamic Ones} 
\author{Mohamed elShehaby, Carleton University
\and
Ashraf Matrawy, Carleton University
\\
\\
\textit{Authors’ draft for soliciting feedback}}




%





\maketitle

\begin{abstract}
Machine Learning (ML) has become pervasive, and its deployment in Network Intrusion Detection Systems (NIDS) is inevitable due to its automated nature and high accuracy compared to traditional models in processing and classifying large volumes of data. However, ML has been found to have several flaws, most importantly, adversarial attacks, which aim to trick ML models into producing faulty predictions. While most adversarial attack research focuses on computer vision datasets, recent studies have explored the suitability of these attacks against ML-based network security entities, especially NIDS, due to the wide difference between different domains regarding the generation of adversarial attacks. To further explore the practicality of adversarial attacks against ML-based NIDS in-depth, this paper presents several key contributions: identifying numerous practicality issues for evasion adversarial attacks on ML-NIDS using an attack tree threat model, introducing a taxonomy of practicality issues associated with adversarial attacks against ML-based NIDS, identifying specific leaf nodes in our attack tree that demonstrate some practicality for real-world implementation and conducting a comprehensive review and exploration of these potentially viable attack approaches, and investigating how the dynamicity of real-world ML models affects evasion adversarial attacks against NIDS. Our experiments indicate that continuous re-training, even without adversarial training, can reduce the effectiveness of adversarial attacks. While adversarial attacks can compromise ML-based NIDSs, our aim is to highlight the significant gap between research and real-world practicality in this domain, which warrants attention.
\end{abstract}


\section{Introduction} 
\label{Introduction}


Machine Learning (ML) applications are present in every corner. ML found its way to network security applications due to its (1) automated nature and (2) ability to process and classify high volumes of data with high accuracy. Thus, ML deployment in Network Intrusion Detection Systems (NIDS) is a foregone conclusion. However, in recent years, researchers highlighted that ML has a lot of drawbacks and flaws \cite{papernot2018sok} \cite{https://doi.org/10.48550/arxiv.2207.07048}, nevertheless, one ML challenge stood out the most in the research community; adversarial 
 attacks. In a nutshell, these adversarial attacks aim to trick ML models, making them produce faulty predictions. There is no doubt that these attacks are a menace to ML. However, it was observed that the majority of adversarial attack research uses computer vision datasets to evaluate its severity \cite{apruzzese2022position}. Thus, recently, a few researchers \cite{apruzzese2022modeling} \cite{merzouk2022investigating} have explored the practicality of adversarial attacks against ML-based network security entities, especially Network Intrusion Detection Systems (NIDS), and implied that attacking them is much more complicated than computer vision.

\noindent \textbf{Contributions:} In this paper, we study, evaluate, and contextualize the practicality of evasion adversarial attacks against NIDS and make several key contributions. \textbf{First}, we identify the questionable practicality prerequisites required to carry out evasion adversarial attacks against NIDS. We introduce an attack tree, a threat modeling technique, to visualize various scenarios attackers might use to launch evasion adversarial attacks against ML-NIDS. The attack tree helps identify leaf nodes with questionable feasibility. \textbf{Second}, we extend the research discussion on the practicality of adversarial attacks against NIDS by introducing a taxonomy of practicality issues associated with adversarial attacks against ML-based NIDS. This taxonomy includes issues reported in the literature, along with our own questions and observations. \textbf{Third}, we identify specific leaf nodes in our attack tree that demonstrate some practicality for real-world implementation and conduct a comprehensive review and exploration of these potentially viable attack approaches. \textbf{Fourth}, our contribution emerged from our observation that the overwhelming bulk of adversarial attack research deals with ML approaches as static models. However, nowadays, numerous ML models are dynamic and constantly evolving. Approaches like stream learning \cite{casas2019should} or online machine learning \cite{fontenla2013online} constantly update their models to make them adaptive to new cases and data (prevent concept drifts). Therefore, we explore and test the effect of continuous training on gradient-based adversarial attacks against NIDS. To the best of our knowledge, these are the first experiments on the impact of continuous training on such invasions against ML-based NIDS. 



\begin{table*}[]
\notsotiny
\centering
\caption{Comparative Analysis of our paper and Related Work on Adversarial Attacks in Network Intrusion Detection Systems ({\footnotesize $\bullet$} = full, \protect\halfcirc{} = partial, $\circ$ = no coverage).}
\label{tab:comparison}
\begin{tabular}{|>{\centering\arraybackslash}m{2.2cm}|>{\centering\arraybackslash}m{1.6cm}|>{\centering\arraybackslash}m{2.5cm}|>{\centering\arraybackslash}m{1.2cm}|>{\centering\arraybackslash}m{1.8cm}|>{\centering\arraybackslash}m{1.2cm}|}
\hline
\textbf{Paper} & 
\textbf{Evasion Adversarial Attacks on NIDS} & 
\textbf{Threat Modeling (Attack Trees) \& Feasibility Assessment$^{\mathrm{*}}$}  & 
\textbf{Practicality Analysis of Attacks} & 
\textbf{Taxonomy of Attack Practicality Issues} & 
\textbf{Impact of Dynamic Learning} \\
\hline
Ennaji et al. \cite{ennaji2025adversarial} & \cellcolor{green!20} {$\bullet$} & \cellcolor{red!20}$\circ$ & \cellcolor{yellow!20} \halfcirc & \cellcolor{red!20}$\circ$ & \cellcolor{red!20}$\circ$ \\
\hline
He et al. \cite{he2023adversarial} & \cellcolor{green!20}{ $\bullet$} & \cellcolor{red!20}$\circ$ & \cellcolor{green!20}{ $\bullet$}  & \cellcolor{red!20}$\circ$ & \cellcolor{red!20}$\circ$ \\
\hline
 Alatwi et al. \cite{alatwi2021adversarial}  & \cellcolor{green!20}{ $\bullet$} & \cellcolor{red!20}$\circ$ & \cellcolor{yellow!20} \halfcirc & \cellcolor{red!20}$\circ$ & \cellcolor{red!20}$\circ$ \\
\hline
 Vitorino et al. \cite{vitorino2023sok} & \cellcolor{green!20}{ $\bullet$} & \cellcolor{red!20}$\circ$ & \cellcolor{green!20}{ $\bullet$} & \cellcolor{red!20}$\circ$ & \cellcolor{red!20}$\circ$ \\
\hline
 Martins et al. \cite{martins2020adversarial}  & \cellcolor{green!20}{ $\bullet$} & \cellcolor{red!20}$\circ$ & \cellcolor{red!20}$\circ$ & \cellcolor{red!20}$\circ$ & \cellcolor{red!20}$\circ$ \\
\hline
 Apruzzese et al. \cite{apruzzese2022modeling}  & \cellcolor{green!20}{ $\bullet$} & \cellcolor{red!20}$\circ$ & \cellcolor{green!20}{ $\bullet$} & \cellcolor{red!20}$\circ$ & \cellcolor{red!20}$\circ$ \\
\hline
 Merzouk et al. \cite{merzouk2022investigating}  & \cellcolor{green!20}{ $\bullet$} & \cellcolor{red!20}$\circ$ & \cellcolor{green!20}{ $\bullet$} & \cellcolor{red!20}$\circ$ & \cellcolor{red!20}$\circ$ \\
\hline
 Alatwi et al. \cite{alatwi2022threat} & \cellcolor{green!20}{ $\bullet$} & \cellcolor{yellow!20} \halfcirc & \cellcolor{red!20}$\circ$ & \cellcolor{red!20}$\circ$ & \cellcolor{red!20}$\circ$ \\
\hline
\textbf{Our Paper (2026)} & \cellcolor{green!20}{ $\bullet$} & \cellcolor{green!20}{ $\bullet$} & \cellcolor{green!20}{ $\bullet$} & \cellcolor{green!20}{ $\bullet$} & \cellcolor{green!20}{ $\bullet$} \\
\hline
\multicolumn{6}{l}{$^{\mathrm{*}}$Note: Feasibility Assessment includes identifying attack tree leaves with questionable and possible feasibilities}
\end{tabular}
\end{table*}

\section{Related Work} \label{Related_Work}
In this section, we review previous related work and classify them into two groups; adversarial attacks against ML-NIDS, and the practicality of adversarial attacks against ML-NIDS.

\subsection{Adversarial attacks against ML-NIDS}
Numerous researchers have examined adversarial attacks against machine learning-based network intrusion detection systems. Alatwi et al. \cite{alatwi2021adversarial} surveyed and categorized the studies of adversarial attacks against ML-based NIDSs and briefly discussed their applicability in real-world scenarios. He et al. \cite{he2023adversarial} has also conducted a comprehensive exploration of white- and black-box adversarial attacks on DNNs, and their potential relevance in NIDS. They also analyzed existing defense mechanisms against these attacks. Furthermore, Martins et al. \cite{martins2020adversarial} examined adversarial machine learning against intrusion and malware detection systems. They concluded that a wide variety of the reviewed attacks could be effective, however, their practicality was not been tested enough in intrusion scenarios. Furthermore, They have stated that some of the datasets in intrusion scenarios are outdated, and most of them are similar and lack variety. Ennaji et al. \cite{ennaji2025adversarial} also surveyed evasion adversarial attacks on NIDS and provided practical recommendations for adapting adversarial strategies to network-specific constraints.

\subsection{Practicality of Adversarial attacks against ML-NIDS}
Some studies have focused on the practicality of adversarial attacks in the network intrusion detection domain; Vitorino et al. \cite{vitorino2023sok} presented an overview of adversarial learning techniques capable of generating realistic examples to deceive NIDSs. Merzouk et al. \cite{merzouk2022investigating} also attempted to inspect the practicality of adversarial attacks to evade network intrusion detection systems. They highlighted some limitations of adversarial attacks against ML-NIDS. However, they have only concentrated on the feature space attacks. Apruzzese et al. \cite{apruzzese2022modeling} aimed to specify and model the abilities and circumstances necessary for attackers to execute practical adversarial attacks on network intrusion detection systems. Both \cite{vitorino2023sok} and \cite{apruzzese2022modeling} primarily reviewed and evaluated attacks from research literature but did not consider the dynamic nature of modern ML models. In a separate study, Apruzzese et al. \cite{apruzzese2022position} emphasized the existence of a gap between researchers and practitioners in the field of adversarial attacks. Their primary focus was not on NIDSs; however, through collaboration between industry and academia, they articulated the general viewpoint that \emph{``Real attackers don't compute gradients.''} In other words, they were unable to identify an attacker who computes gradients to craft real-world adversarial attacks.


\subsection{Comparison with Related Work}

As seen in Table \ref{tab:comparison}, which shows the comparative analysis of our paper and related work on adversarial attacks in the NIDS domain, our work, unlike other relevant related work, covers all the comparison points in the list. To the best of our knowledge, none of the previous work has explored and tested the practicality of adversarial attacks against dynamic ML models in the intrusion detection domain. Moreover, to the best of our knowledge, none of the previous work has merged systematic literature review, threat modeling techniques, and testing with evasion adversarial attacks on NIDS to evaluate the practicality of evasion adversarial attacks in the NIDS domain. Alatwi et al. \cite{alatwi2022threat} conducted threat modeling techniques for adversarial attacks against NIDS; however, their attack tree covered all adversarial attacks, while ours focuses on just evasion attacks, so naturally ours was more in-depth, covered more angles, and identified attack tree leaves with questionable and possible feasibilities. Unlike their work, ours was backed by systematic literature review and testing.

\section{Methodology of Systematic Literature Review}
\label{Methodology}

\begin{figure}[]
    \centering
    \includegraphics[width=0.7\linewidth,keepaspectratio=true]{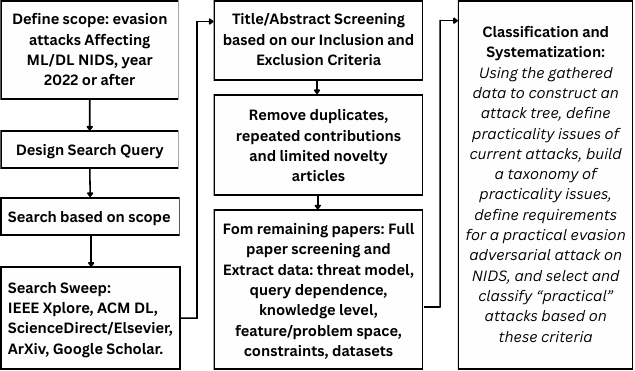}
    \caption{Systematic Literature Review}
    \label{fig:sok}
\end{figure}

We conduct a systematic literature review following a six-phase approach as shown in Figure \ref{fig:sok}. Our methodology begins by defining the scope to focus on evasion attacks affecting ML/DL NIDS published in 2022 or after. We design a systematic search query using three search terms in conjunction with a temporal constraint (Table~\ref{tab:search_query}): adversarial attack terms, network intrusion detection terms, and machine learning terms. The search is conducted across IEEE Xplore, ACM Digital Library, ScienceDirect/Elsevier, ArXiv, and Google Scholar. Following our initial title and abstract screening using the criteria in Table \ref{tab:inclusion_exclusion}, we conduct a second, thorough screening to remove duplicates, repeated contributions, and articles with limited novelty. We then perform full-text screening and extract key data including threat models, knowledge requirements, and implementation constraints. Finally, we use the gathered data to construct an attack tree, identify practicality issues in current attacks, define requirements for practical adversarial attacks on NIDS, and systematically classify papers that meet our practicality criteria.


The temporal constraint of focusing on publications from 2022 onwards is motivated by He et al.'s \cite{he2023adversarial} comprehensive survey in \textit{IEEE Communications Surveys \& Tutorials}, which stated that ``existing adversarial attacks lack a comprehensive evaluation of the maliciousness of the adversarial examples'' and highlighted insufficient practical implementation considerations in adversarial attacks against NIDS. This critical assessment of pre-2022 research motivates our investigation: we aim to determine whether adversarial attacks published from 2022 onwards have become more practical and realistic in addressing real-world deployment constraints, or if these fundamental limitations persist in recent literature.

\begin{table}[]
\centering
\scriptsize
\caption{Search query format used to retrieve papers. The query is composed of three search terms in conjunction with a temporal constraint. The values in the second column replace each search term.}
\label{tab:search_query}
\begin{tabular}{|p{3cm}|p{7cm}|}
\hline
\textbf{Search Query} & \textbf{<search\_term\_1> AND <search\_term\_2> AND <search\_term\_3> AND <temporal\_constraint>} \\
\hline
\textbf{<search\_term\_1>} & ``adversarial attack'' OR ``adversarial example'' OR ``adversarial perturbation'' OR ``adversarial machine learning'' AND ``evasion''\\
\hline
\textbf{<search\_term\_2>} & ``network intrusion detection'' OR ``intrusion detection system'' OR ``NIDS'' OR ``IDS''  \\
\hline
\textbf{<search\_term\_3>} & ``machine learning'' OR ``ML'' \\
\hline
\textbf{<temporal\_constraint>} & Publication year: 2022 or after \\
\hline
\end{tabular}
\end{table}

\begin{table}[]
\centering
\footnotesize
\caption{Inclusion and Exclusion Criteria for Systematic Literature Review}
\label{tab:inclusion_exclusion}
\renewcommand{\arraystretch}{1.4} 
\begin{tabular}{@{} l p{10cm} @{}}
\toprule
\textbf{Criteria Type} & \textbf{Description} \\
\midrule
\textbf{Inclusion Criteria} & 
\textbullet\ Published between 2022 and 2026 \newline
\textbullet\ Focuses on evasion adversarial attacks \newline
\textbullet\ Targets ML/DL-based network intrusion detection systems \newline
\textbullet\ Written in English \\
\midrule
\textbf{Exclusion Criteria} & 
\textbullet\ Targets defenses, not evasion adversarial attacks \newline
\textbullet\ Focuses on poisoning attacks, model stealing, or backdoor attacks only \newline
\textbullet\ Targets domains other than network intrusion detection \newline
\textbullet\ Insufficient technical detail for analysis \newline
\textbullet\ Duplicate publications (only most complete version retained) \\
\bottomrule
\end{tabular}
\end{table}






\section{Background} \label{Background}

\subsection{ML-based NIDS}
\begin{figure}[]
    \centering
    \includegraphics[width=0.85\linewidth,keepaspectratio=true]{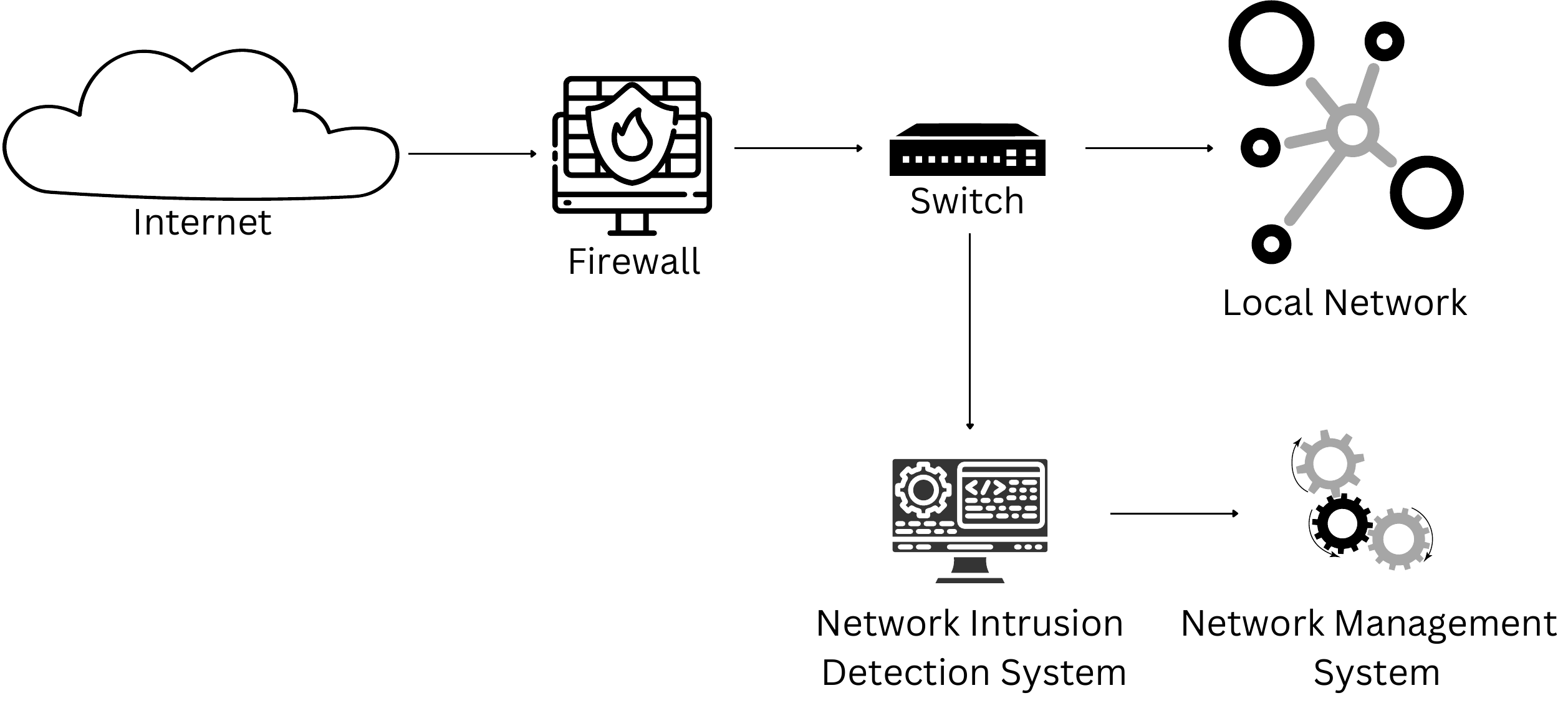}
    \caption{The Deployment of Network Intrusion Detection System}
    \label{fig:NIDS}
\end{figure}

Network intrusion detection systems (NIDS) are tools designed to identify threats against a network. They send alerts to network management in case of breaches. Figure  \ref{fig:NIDS} shows the deployment of Network Intrusion Detection Systems. There are two main NIDS techniques: signature-based and anomaly-based. Signature-based detection identifies threats by detecting predetermined attack patterns (signatures). This type of detection has a low false positive rate in an environment of well-defined attacks. However, its main disadvantage is that it cannot detect newer or unknown attacks \cite{kumar2012signature}. Anomaly-based detection, on the other hand, identifies attacks by detecting deviations from normal activity. This approach makes it able to detect unknown attacks. However, because of the difficulty of defining the boundary between normal and abnormal behavior, anomaly-based detection can have a high false positive rate \cite{chandola2009anomaly}. Therefore, to solve the accuracy issues in anomaly-based detection, researchers have explored the use of machine learning approaches (like Support Vector Machine (SVM), Decision Tree (DT), Artificial Neural Networks (ANN), etc.) \cite{ahmad2021network} and found them very accurate in classifying large amounts of data. However, some researchers highlighted that machine learning can be vulnerable to adversarial attacks \cite{papernot2018sok}. We will further explore these attacks later in this section.

\subsection{Dynamic Machine Learning}

Concept drift in machine learning is the unforeseeable statistical changes in the prediction target's properties \cite{lu2018learning}. In other words, the norm changes over time, so the ML model makes wrong predictions. In an adversary environment, where attackers change their techniques all the time, like the network/cybersecurity domain, concept drifts will more than likely occur \cite{casas2019should}. Thus, to counter these drifts ML models must be dynamic, i.e, continuously evolving. Numerous terms are used to describe dynamic ML models (e.g., Online ML, Evolving ML, Continuous/Continual ML, and Stream ML). In the ML community, the implementation and definition of these kinds of ML techniques could differ from one study to another. However, the following concept is common; \emph{``The initial training set loses its validity as time passes due to changes of conditions in the aimed task. Thus, a mechanism to update a given model in order to adapt to new conditions is necessary''} \cite{fontenla2013online}. In simple terms, dynamic ML models must be updated and re-trained all the time to resist the effect of concept drift. There are different techniques for re-training (or continuous training); for example, adaptive windowing, where the model is re-trained when a concept drift is detected \cite{bifet2007learning}. However, in practice, there are more elementary solutions, with less overhead on the system, like periodically re-training the model with recent data. In our tests, we use the latter approach.

\subsection{Adversarial Attacks Against Network Security and Defenses}


\textbf{Adversarial examples}, first discovered in 2013 by Szegedy et al. \cite{szegedy2013intriguing}, are inputs that cause machine learning models to make incorrect predictions despite being very similar to legitimate examples. Soon after, Goodfellow et al. \cite{goodfellow2014explaining} introduced the \textbf{FGSM (Fast Gradient Sign Method)} attack, which creates adversarial examples by adding small \textbf{perturbations} equal to the sign of the elements of the gradient of the cost function. These attacks have become more sophisticated and diverse. In the next subsections, we will explore their different types and categories.

\subsubsection{Taxonomy of Adversarial Attacks in Network Security}

\textbf{Attackers' Approaches:}
In \textbf{Evasion Attacks}, which is the main focus of this paper, attackers aim to deceive the ML model during the decision-making process by adding carefully crafted perturbations to the input data. Like \cite{ibitoye2019analyzing}, where a perturbation is added to the features to make the IoT ML intrusion detection model output the wrong classification. As for \textbf{Poisoning Attacks}, the attackers add malicious samples to the training datasets (dataset contamination). Like \cite{nguyen2020poisoning}, where a data poisoning attack against a federated learning-based IoT intrusion detection system allowed the adversaries to implant a backdoor into the ML detection model to wrongly classify malicious traffic as benign. 
In \textbf{Model Stealing Attacks}, an attacker can probe a black box model to extract all the vital information to create a ``clone'' model very close to the target model. It is worth mentioning that ML models in a monumental system like NIDS are very expensive, and requires a lot of resources to create. Thus, stealing them will be a massive loss to the associations that created them. The main focus of this paper is evasion attacks. Regarding \textbf{Backdoor Attacks}, the attackers aim to embed a backdoor into the ML model. The backdoored model operates normally when processing inputs lacking the trigger. However, it is misdirected to perform the attacker’s sub-task once the secret trigger is presented in the input \cite{gao2020backdoor}.

\begin{figure*}[]
    \centering
    \includegraphics[width=0.7\linewidth,keepaspectratio=true]{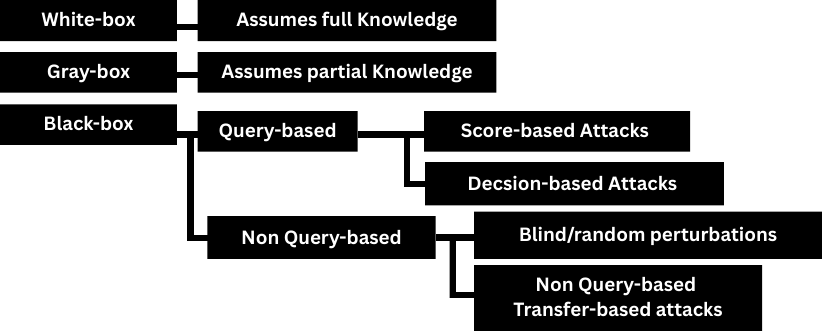}
    \caption{Attackers' Knowledge Types}
    \label{fig:Box}
\end{figure*}


\textbf{Attackers' Knowledge:} As seen in Figure \ref{fig:Box}, there are three types of knowledge in adversarial attacks: white-, gray-, and black-box attacks. In \textbf{White-box Attacks}, the attackers have complete knowledge of the attacked ML model (which ML algorithm was used, gradient, hyperparameters, model architecture, parameters, etc.). The complete opposite is \textbf{Black-box Attacks}, where the attackers have no knowledge of the attacked ML model's internals. There are query-based and non-query-based black-box attacks. The query-based attacks can be score-based, where the attacker queries the model to obtain the confidence score to get an approximation to the decision boundary of the target model, or decision-based, where attackers only have access to the final model decision (e.g., class label) without any confidence scores or probability information to create a surrogate model where they can then attack with white-box attack methods. Query-based attacks present significant practicality limitations as outsiders don't have access to NIDS outputs, so they cannot get confidence scores nor decisions. Non-query-based black-box attacks, the most practical attacks, can be blind or random non-gradient perturbations to input network traffic, which produce inconsistent results, or non-query-based transfer-based attacks, where an attacker trains a surrogate model on the same data distribution as the target model to launch black-box attacks using the transferability property of ML models. \textbf{Gray-box Attacks} lie between white-box and black-box scenarios, where attackers have partial knowledge of the attacked ML system.

\textbf{Attack's Space:}
\begin{figure*}[]
\centering
\includegraphics[width=1\linewidth,keepaspectratio=true]{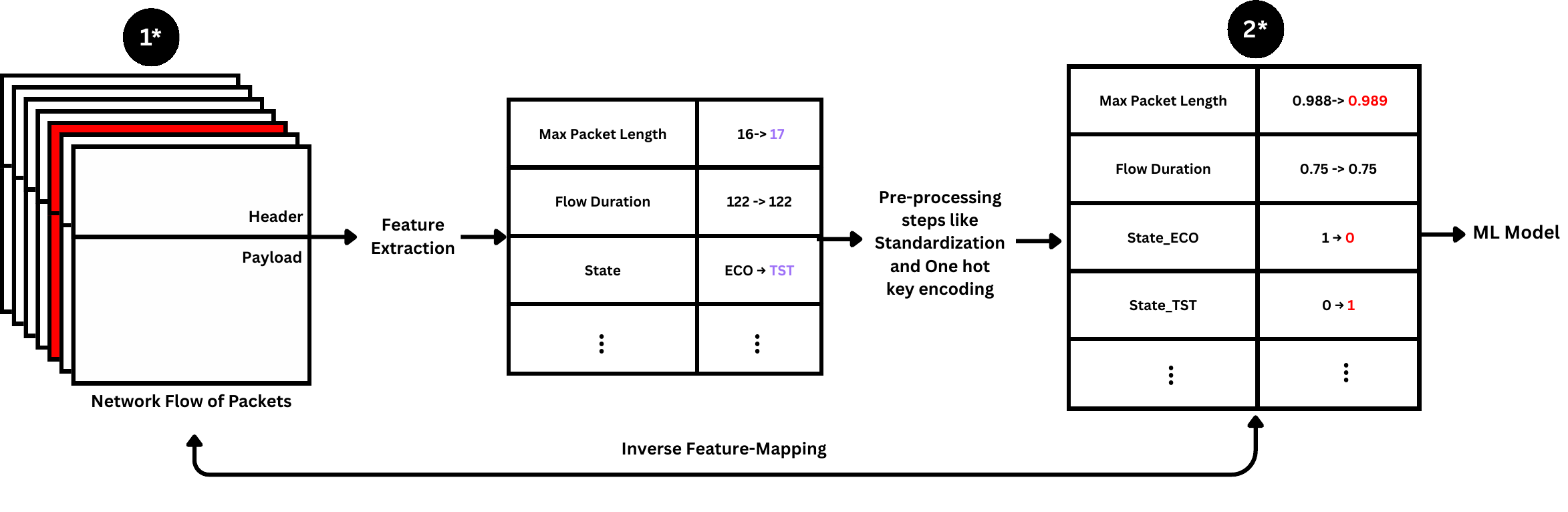}

	\caption{  \setlength{\baselineskip}{12pt} Feature-Space vs. Problem-Space Perturbations in NIDS Domain: If modifications occur at \protect\circled{1*} to the input network flow, this constitutes a problem-space attack; if perturbations occur at  \protect\circled{2*} to the input feature vector, this constitutes a feature-space attack. In problem-space attacks, modifications such as increasing the size of one packet (the red one, for example) to alter the Max Packet Length feature must be mapped to the ML model's input feature vector. This process is called inverse feature mapping \cite{pierazzi2020intriguing}. Note that the perturbations must remain malicious after both feature extraction and pre-processing (red perturbations), not just after feature extraction alone (purple perturbations).} 
	\label{fig:ProbvsFeat}
	\centering
\end{figure*}
 Ibitoye et al. \cite{ibitoye2025threat} presented the space term in adversarial attacks taxonomy in the network security domain. In \textbf{Feature-space} attacks, the attacker only modifies (by adding perturbations) the ML model input feature vector to craft the attack. As for the \textbf{Problem-space} attacks, the attacker modifies the actual file (malware, pcap, PE (Portable Executable), etc.) to trick the attacked model. Figure \ref{fig:ProbvsFeat} shows the difference between feature-space and problem-space Perturbations in the NIDS domain.

 
\subsubsection{Defenses Against Adversarial Attacks}

\begin{figure}[]
    \centering
    \includegraphics[width=0.77\linewidth,keepaspectratio=true]{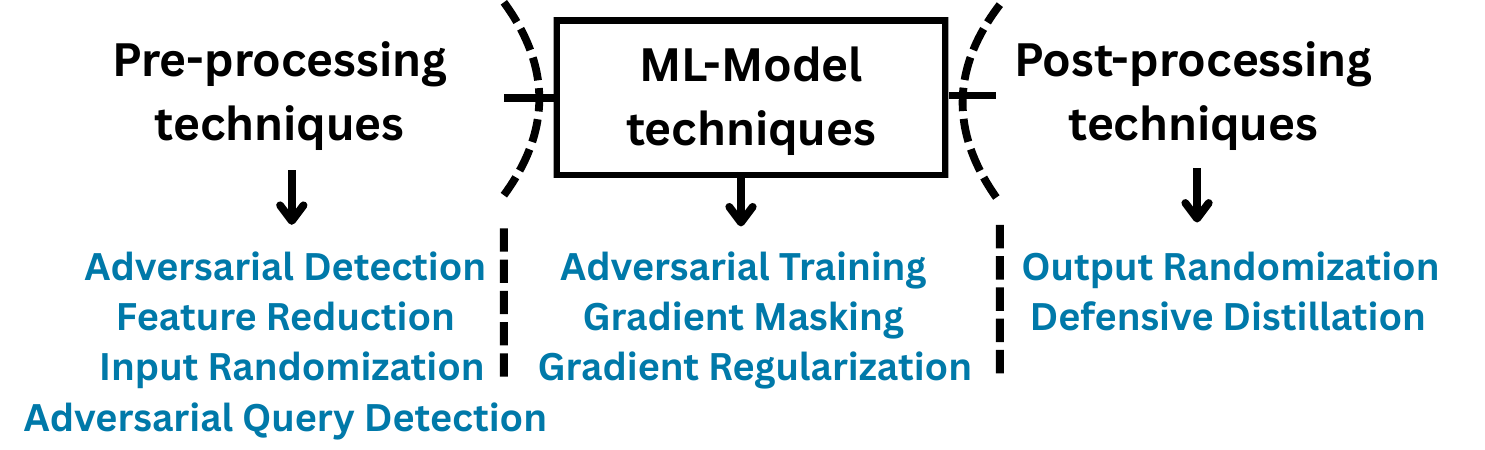}
    \caption{Defenses Against Adversarial Attacks}
    \label{fig:Def}
\end{figure}

There are multiple approaches to defending against adversarial attacks; as seen in Figure \ref{fig:Def} we classify them into 3 classes: pre-processing techniques, ML-model hardening techniques, and post-processing techniques. \textbf{(1) Pre-processing techniques} include defense methods preceding ML model inference like \textbf{Adversarial Detection}; for example, Ren et al. \cite{ren2022towards} proposed adversarial attack detection methods by using the causal inference technique. He et al. \cite{he2023adversarial} classified Adversarial Detection techniques into secondary classifier, projection-based, statistics-based, and mutation-based, which is the most practical because it is applicable and generalizable \cite{he2023adversarial}. \textbf{Feature Reduction} might decrease the effectiveness of perturbations, thereby making the ML classifier less vulnerable to evasion attacks as it reduces the attack-space. Elshehaby et al. \cite{elshehaby2026novel} took it further by applying domain-specific constraints rather than blindly or randomly reducing the feature vector. They used their novel score to select resilient features during feature selection or mask non-resilient features. \textbf{Input Randomization} involves adding random padding on inputs to lessen the effect of adversarial perturbations \cite{ibitoye2025threat}. \textbf{Adversarial Query Detection} aims to defend against query-based black-box adversarial attacks by detecting these queries. Usually, these defenses analyze query patterns and sequences to identify suspicious behavior in queries received by the system \cite{rashid2023malprotect}. \textbf{(2) ML-model hardening techniques} involve defense methods that make the ML model more robust against evasion adversarial attacks; these methods are implemented during model development or training like \textbf{Adversarial Training} which involves inserting adversarial examples into the training dataset to enhance the model's robustness against such attacks \cite{abou2020evaluation} \cite{abou2020investigating}. \textbf{Gradient Masking} modifies machine learning models in an attempt to obscure their gradient from an attacker \cite{nayebi2017biologically} since most evasion adversarial attacks are based on the use of gradient. \textbf{Gradient regularization} is a technique that constrains gradient magnitudes across neural network layers to improve model robustness and prevent overfitting. This approach has been studied as a defense against adversarial attacks, with researchers like Ross et al. \cite{ross2018improving} demonstrating that gradient regularization enhances deep neural network resilience to adversarial perturbations. \textbf{(3) Post-processing techniques} involve defense methods post ML model inference and development. \textbf{Output randomization} is a defense strategy that exploits the limited knowledge available to black-box attackers, who can only access model confidence scores without knowing the underlying architecture. Kwon et al. \cite{kwon2020advguard} developed AdvGuard, which injects noise into the softmax layer to randomize confidence values returned to queries. This randomization disrupts black-box attacks that depend on consistent confidence feedback, making adversarial example generation infeasible by providing unreliable scoring information. \textbf{Knowledge distillation}, initially developed by Hinton et al. \cite{hinton2015distilling} for compressing large neural networks into smaller ones, was later repurposed by Papernot et al. \cite{papernot2016distillation} as an adversarial defense mechanism. This adapted approach uses the original network's outputs as training targets for a distilled model to reduce overfitting and increase the robustness of the model. Similar to ensemble learning, researchers have proposed combining multiple defense strategies to enhance robustness against adversarial attacks \textbf{(Ensemble Defenses)}. Adaptive Continuous Adversarial Training (ACAT) \cite{11160917}\cite{SHE024introducing} exemplifies this approach by merging adversarial training with detection, incorporating newly detected adversarial examples into ongoing training processes.

While these defense mechanisms might provide protection against adversarial attacks, they suffer from significant limitations. The most critical drawback is the substantial system complexity and associated costs, requiring additional computational resources, specialized expertise, and ongoing maintenance overhead. Pre-processing techniques like adversarial detection and input randomization introduce latency, computational burden, and may degrade model performance on legitimate traffic. ML-model hardening approaches such as adversarial training dramatically increase training time and computational requirements, often reducing model accuracy. Post-processing defenses add another layer of computational complexity and may introduce inconsistent outputs affecting system reliability.

\section{Identifying Infeasible and Feasible Adversarial Attack Prerequisites Through an Attack Tree Threat Model}
\label{tree}

\begin{figure}[]
\captionsetup[subfloat]{labelfont=scriptsize,textfont=scriptsize}
\subfloat[{\scriptsize Adversarial Example Generation in the Computer Vision Domain}]{%
  
\centering
\includegraphics[width=0.93\linewidth,keepaspectratio=true]{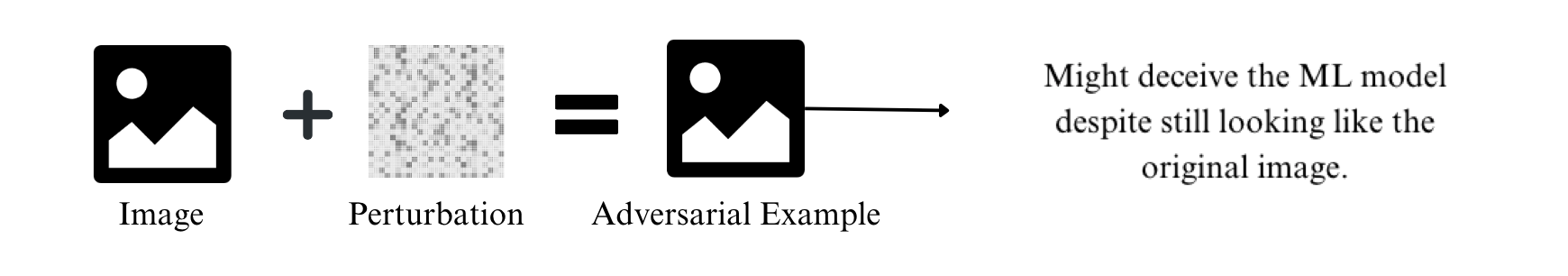}

    \label{fig:ImagePert}
}

\subfloat[{\scriptsize Adversarial Example Generation in the Network Security Domain}]{%

  \centering
\includegraphics[width=0.93\linewidth,keepaspectratio=true]{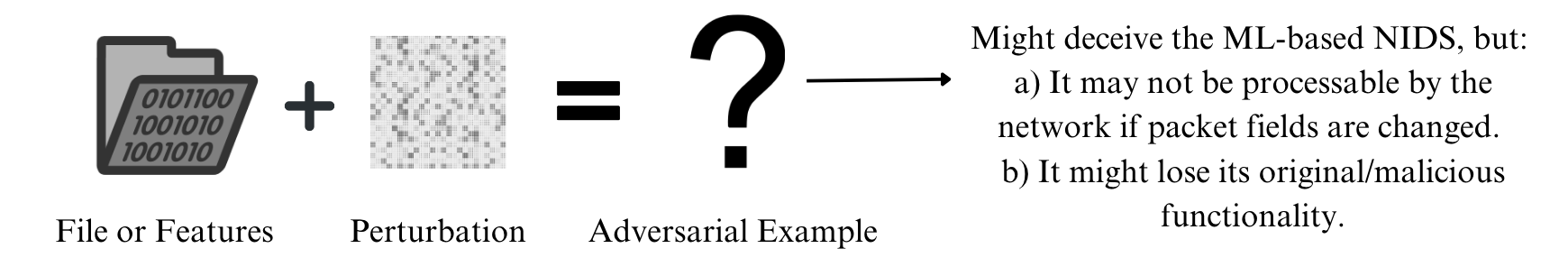}

  \label{fig:FilePert}
}
\caption{Adversarial Examples Generation}
 \label{fig:Adversarial_Examples}
\end{figure}

The majority of the adversarial attacks were designed and tested in the literature using computer vision datasets \cite{apruzzese2022position}. \textbf{Because of the focus on the computer vision datasets, it is not clear if these attacks will be applicable in the same way to the network/cybersecurity domain which has more adversaries and more security constraints and measures, making it much more complicated to evade.} This contrast is depicted in Figure \ref{fig:Adversarial_Examples}, which illustrates the crafting of evasion adversarial attacks in both the computer vision and network domains. Thus, in this section, we present an attack tree, depicted in Figure \ref{fig:ATree}, that showcases the various strategies attackers might employ to execute evasion adversarial attacks. Additionally, we identify leaf nodes (prerequisites) with questionable feasibility marked by the '?' sign, and red cells denote leaf nodes with some feasibility. OR nodes represent disjunctions between alternative approaches, while conjunctions indicate required simultaneous conditions.

The attack tree presented in Figure \ref{fig:ATree} outlines an Adversarial Evasion Attack strategy targeting Machine Learning-based Network Intrusion Detection Systems (ML-NIDS). The tree diverges into two main branches, representing the two alternative ways attackers can fool the model: Adversarial Features (Feature space) and Adversarial Packets (Problem-Space). The right branch focuses on the manipulation of feature vectors to deceive ML-NIDS. To achieve this, attackers require access to the feature vector and the ability to introduce perturbations to specific features. However, directly accessing the feature vector presents a significant obstacle for external attackers, making this approach highly impractical. Introducing perturbations to the feature vector requires identifying perturbable features (not all features are susceptible to manipulation, as will be discussed later) and requires up-to-date knowledge of the underlying ML-NIDS model. Identifying perturbable features requires information about the specific features employed by the target ML-NIDS. Acquiring this information through insider knowledge is highly unlikely. An alternative option for attackers is to make informed assumptions or predictions regarding certain features, an approach fraught with uncertainties. As for up-to-date knowledge of the model, the attacker could obtain it through insider knowledge (questionable practicality) or by querying the model directly. However, as illustrated in Figure \ref{fig:NIDS}, accessing the model's input and output is very improbable for an outsider, as NIDSs are not designed to provide feedback when queried \cite{debicha2023adv}. Nonetheless, side-channel querying represents another potential avenue. In essence, the fundamental limitation of directly accessing the feature vector renders the right branch (feature-space attacks) highly impractical for real-world scenarios.

In the left branch of the tree, we delve into problem-space attacks, which involve manipulating network packets to evade ML-NIDS detection. In this scenario, the attacker needs to gain access to network packets and introduce perturbations to them. Accessing network packets is comparatively more feasible than accessing the feature vector of the model, making problem-space attacks appear more practical than feature-space attacks. However, adding perturbations to network packets requires the attacker to preserve network functionality, maintain malicious functionality (since it's an attack), handle side-effect features (a.k.a. collateral damage features) \cite{pierazzi2020intriguing} \cite{elshehaby2026novel}, and reverse engineer the packets to produce perturbed features after extraction, in addition to addressing all the constraints presented in the right branch (feature-space attacks), since the main aim of problem-space attacks is to add perturbations to features after feature extraction. The tree shows that maintaining network functionality and malicious functionality are required simultaneously (AND condition), while handling side-effect features represents an additional constraint that attackers must address when modifying network packets. The reverse engineering process presents multiple pathways with their own constraints. We find that maintaining the malicious functionality of network packets presents uncertain feasibility, as existing adversarial attacks against ML-NIDS lack a comprehensive evaluation of the maliciousness of their generated examples. These attacks primarily focus on evading detection by the model, neglecting the preservation of the intended malicious behavior \cite{he2023adversarial}. Additionally, the process of reverse engineering the packets entails similar requirements as the feature-space attack outlined in the right branch of the tree, including the challenging aspects of obtaining model knowledge and identifying perturbable features.

The attack tree illustrates that evasion adversarial attacks against ML-NIDS remain highly challenging. Although problem-space attacks may appear more practical than feature-space approaches, both attack vectors face significant real-world constraints. However, the red branches in our attack tree indicate certain aspects that might possess some feasibility, which we will explore in subsequent sections to provide a comprehensive assessment of the attack landscape.

\begin{figure*}
\hspace*{-2.2cm}
\centering
{\tiny
\begin{forest}
for tree={
    draw, 
     text centered 
    }
[ Adversarial\\Evasion Attack\\Against ML-NIDS
  [Create Adversarial\\Packets that produces\\ perturbed features\\(Problem-Space), angle below [Access Network\\Packets ,  name =y] [Add perturbations\\on Network Packets, angle below [Maintain\\Network\\Functionality] [Maintain\\Malicious\\Functionality] [Reverse Engineer\\the packets in\\a way that\\produces perturbed\\ features after\\the features\\extraction, angle below
  [Handle\\Side-effect\\(Collateral\\Damage)\\Features][OR Node [Know\\up-to-date\\info\\on training\\data of the\\target model\\to Create\\surrogate\\model, fill=pink][Add random\\or Blind\\Perturbations, fill=pink][Know\\up-to-date\\information of\\the model,  name =x [Insider\\ Knowledge\\(White-Box)\\?] [Obtain\\Model's\\input and\\output\\(Black-Box) ,angle below [Query\\the model\\?][Side-\\Channel\\Querying, fill=pink]]]][Pick perturbable\\ Features[Predict some\\Feature Vector Items\\and their pre-processing] [Insider\\Knowledge\\(White-Box)\\?]]
  ]]]
  [Adversarial\\Features\\(Feature-Space), angle below [Access Feature\\Vector\\?] [Add perturbations\\on Perturbable\\Features, angle below
  [Pick perturbable\\ Features[Predict some\\Feature Vector Items\\and their pre-processing] [Insider\\Knowledge\\(White-Box)\\?]][OR Node [Know\\up-to-date\\info\\on training\\data of the\\target model\\to Create\\surrogate\\mode][Add random\\or Blind\\Perturbations][Know\\up-to-date\\information of\\the model,  name =xx [Insider\\ Knowledge\\(White-Box)\\?] [Obtain\\Model's\\input and\\output\\(Black-Box) ,angle below [Query\\the model\\?][Side-\\Channel\\Querying]]]]]]]
\end{forest}
}

\caption{Attack tree for adversarial evasion attacks against ML-NIDS. {\large<} indicates a disjunction (OR), \textbf{{\large$\sphericalangle$}} indicates a conjunction (AND), \textbf{?} denotes a leaf node with uncertain feasibility (questionable practicality), and red cells denote leaf nodes with some feasibility for later exploration.}
\label{fig:ATree}

\end{figure*}

\section{Taxonomy of Practicality Isues of Adversarial Attacks Against ML-NIDS} \label{taxo}

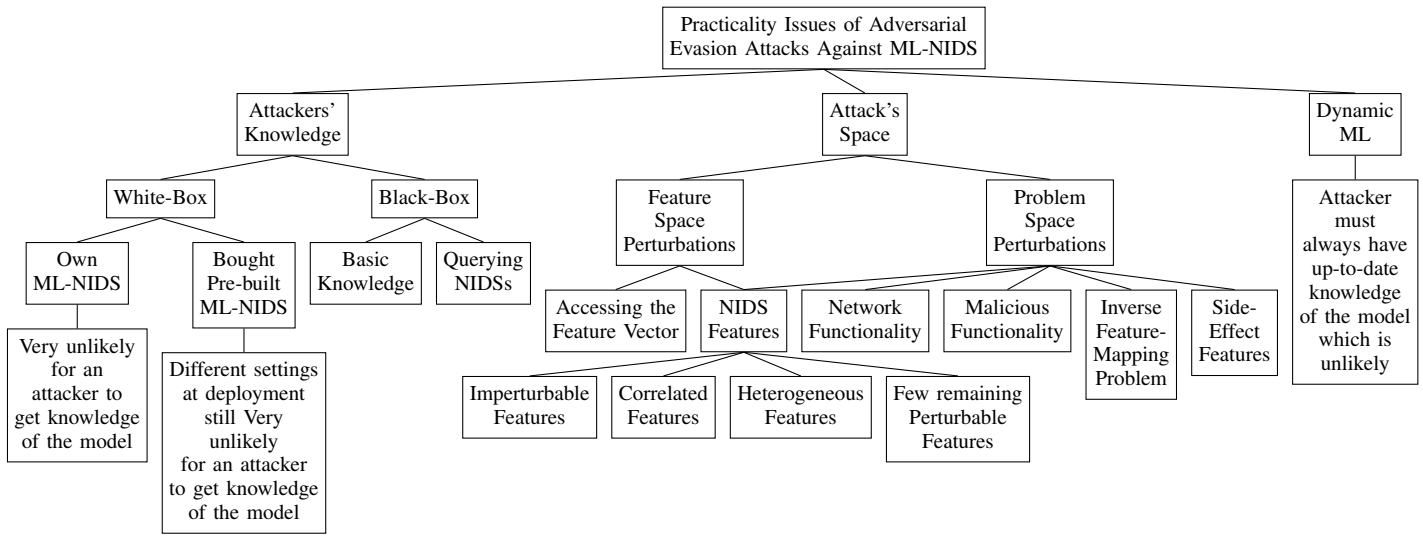
\begin{figure*}
\hspace*{-2.6cm}
{\tiny
\begin{forest}
for tree={
    draw, 
     text centered 
    }
[ Practicality Issues of Adversarial\\Evasion Attacks Against ML-NIDS
  [Attackers'\\Knowledge[White-Box [Own\\ML-NIDS [Very unlikely\\for an\\attacker to\\get knowledge\\of the model]] [Bought\\Pre-built\\ML-NIDS [Different settings\\at deployment\\still Very\\unlikely\\ for an attacker\\ to get knowledge\\of the model]]] [Black-Box [Preliminary\\Knowledge][Querying\\NIDSs]]]
  [Attack's\\Space [ Feature\\Space\\Perturbations [Accessing the\\Feature Vector] [NIDS\\Features ,name=x [Imperturbable\\Features] [Correlated\\Features] [Heterogeneous\\Features] [Few remaining\\Perturbable\\Features]]]
  [ Problem\\Space \\Perturbations, name =y [Network\\Functionality][Malicious \\Functionality][Inverse\\Feature-\\Mapping\\Problem] [Side-\\Effect\\Features]]]
  [Dynamic\\ML [Attacker\\must\\always have\\up-to-date\\knowledge\\of the model\\which is\\unlikely]] 
]
{ 
\draw[-] (y.south) to node[midway,above]{{\footnotesize \textit{}}} (x.north);
}
\end{forest}
}

\caption{Taxonomy of Practicality Isues of Adversarial Attacks Against ML-NIDS, Directed Acyclic Graph (DAG)}
\label{fig:Practicality}
\end{figure*}

In this section, we explore the issues and challenges of adversarial attacks against ML-NIDS. Our taxonomy, illustrated in Figure \ref{fig:Practicality}, highlights the gap between research and real-world scenarios. It offers a different perspective on the practicality of adversarial evasion attacks against NIDS compared to the attack tree by summarizing most reported practicality issues from the literature and our notes within a single Directed Acyclic Graph (DAG). This taxonomy aims to bridge the gap between research and real-world scenarios.



\subsection{Attackers' Knowledge}

\subsubsection{White-Box Feasibility}
Organizations with dedicated security teams may possess the expertise and budget to develop their own Network Intrusion Detection Systems (NIDS). Alternatively, other organizations may choose to purchase pre-built solutions. In the former scenario, acquiring knowledge of a private model is considered highly improbable for an attacker, as white-box attacks require extensive information such as the ML model utilized, gradients, feature set, and more. Even with publicly available pre-purchased models, the diverse deployment settings make obtaining accurate knowledge challenging for an attacker.
\subsubsection{Black-Box Feasibility}
As mentioned in section \ref{Background}, attackers in black-box settings have no knowledge of the target ML model. There is no doubt that black-box evasions are the more practical type of adversarial attack \cite{alatwi2021adversarial} as obtaining all the needed information about the target model is extremely difficult in the network security domain. In the next sub-section, we will explore the feasibility of black-box attacks against ML-based NIDS. 
 \begin{itemize}
\item \textbf{Preliminary Knowledge:} Even under black-box assumptions, where the attacker lacks knowledge of the model itself (e.g., architecture, parameters, gradients) and relies on transfer-based attacks or random perturbations, some preliminary knowledge remains necessary in the NIDS domain, particularly regarding pre-processing, feature extraction, and feature selection. As shown in Figure \ref{fig:ImagePert}, in computer vision, adversarial examples are generated by slightly modifying pixel values, and the medium is well understood because attackers perturb pixels in a digital image (or add physical perturbations to physical objects, such as a stop sign). By contrast, in the problem-space of network intrusion, the medium to be perturbed is often unknown to the attacker. ML-based NIDSs employ distinct pre-processing pipelines and feature engineering choices. Without awareness of the target model’s selected features and pre-processing methods, attackers do not know what to perturb or how to perturb it. \textbf{As shown in Figure \ref{fig:ProbvsFeat}, problem-space perturbations must produce corresponding feature-space perturbations after pre-processing on the selected features; thus, knowledge of these aspects is essential.} Moreover, many works present black-box attacks while assuming knowledge of the target’s feature set, feature extractor, or feature transformers (e.g., Kuppa et al. \cite{kuppa2019black}). In our view, when attackers possess such detailed information about feature management and selection, the attack no longer qualifies as black-box.

However, as stated in our attack tree, some attackers might attempt to ``predict some feature vector items'' by assuming that certain features are selected based on research and assessment of commonly used and popular NIDS domain datasets. While this approach might be practical in some scenarios, there is no guarantee that the target ML-NIDS actually uses those predicted features. Furthermore, ElShehaby et al. \cite{elshehaby2026novel} introduced a defense mechanism that specifically counters this approach by selecting only features that are not perturbable or hard to perturb in the problem space using their Perturb-ability Score (PS), which might render the assumption of ``Predict some feature vector items'' inapplicable.

\item \textbf{Querying NIDS:} As mentioned in section \ref{Background}, one of the main corners of black-box adversarial attacks is querying the target model to obtain a substitute model. Then using that alternative model, the attackers can get all the knowledge they need to craft their adversarial attack. It would be reasonable to assume that only the system management teams have access to the output of NIDS, as seen in Figure \ref{fig:NIDS}. Thus, it is not clear how querying NIDSs in black-box attack settings is possible if the attacker is not an insider. Debicha et al. clearly stated that NIDSs are not designed to provide feedback \cite{debicha2023adv}. 

However, some threat models in adversarial attack research assume a powerful attacker might gain access to the NIDS system and control it to obtain input-output pairs for querying. This assumption raises a critical question: if an attacker possesses such extensive system access and control, why would they need to craft complex gradient-based evasion attacks when they could simply employ straightforward trial-and-error approaches? With direct access to the system, they could systematically test various perturbations in all directions, observe the system's responses, and iteratively refine their attacks based on what proves effective, eliminating the need for sophisticated gradient computations or complex mathematical optimization techniques. We argue that if the attacker possesses that much power, then it is no longer an ML/adversarial attack problem anymore.

\end{itemize}


\subsection{Attack's Space}

\subsubsection{Feature-Space Perturbations}
In feature-space attacks, the attacker just changes the ML model input feature vector to fool the target model by adding perturbations to the features.

\textbf{Accessing the Feature Vector:} Normally, attackers don't have access to the feature vector after the feature extraction phase in the ML-based NIDS pipeline. So how would feature-space adversarial attacks against NIDS be considered practical? That is an important question to answer considering that most of NIDS's adversarial attack research is in feature-space attacks \cite{apruzzese2022modeling}.

 \textbf{NIDS Features:} As previously mentioned, knowing which features, and feature transformation techniques are used by the target model is a very hard task for the attacker. However, even if the attackers know the selected features and have access to them, adding perturbations to the ML-based NIDS features is a complicated task \cite{merzouk2022investigating}. Unlike the computer vision domain, numerous constraints make crafting adversarial attacks in the NIDS space tough. 
 \begin{itemize}

     \item  \textbf{Imperturbable/Hard-to-Perturb Features:} A sizable portion of features in the NIDS datasets are boolean (such as flags) and discrete numbers (such as the protocol feature). The presence of these features doesn't give flexibility to the attackers to add perturbations that maximize the loss to fool the target model. For example, the attacker cannot add a perturbation to a flag in the positive direction if its value is 1 (because the maximum value is 1). The same goes for a discrete feature; for instance, the attacker cannot add an arbitrary value to the protocol feature because some protocol numbers are unassigned. Moreover, some numerical values have similar limitations; for example, the attacker cannot change the packet length feature to a negative value. Furthermore, perturbing some features in problem-space can totally change the functionality of the network flow, which can make the flow lose its malicious or network functionality, such as altering the destination IP \cite{elshehaby2026novel}. Moreover, attackers cannot access some features such as backward features (from destination to source)  \cite{elshehaby2026novel}, which represent the target system's response characteristics including backward packet count, backward bytes, and backward inter-arrival times, as these are determined by the receiving system's behavior rather than the attacker's input.

     \item \textbf{Correlated Features:} NIDS typically extracts various features from network traffic to identify malicious activity. These features are not always independent and usually exhibit dependencies or correlations among them. For example, TCP flags in packets and TCP as the transport protocol are correlated \cite{sheatsley2022adversarial}. These correlations constrain attackers; for example, an attacker cannot simultaneously increase the flow duration while decreasing both the forward and backward inter-arrival times \cite{elshehaby2026novel}. Unlike in computer vision, in the NIDS domain, attackers cannot simply add perturbations to independent pixels; instead, they must consider the complex relationships between different features, which complicates the crafting of attacks in the NIDS domain.

     \item \textbf{Heterogeneous Features:} In the NIDS domain, features extracted from network traffic fall into different categories with varying types and characteristics. This diversity poses challenges for attackers crafting adversarial attacks. Unlike computer vision, where the perturbed medium is of one type, the features in the NIDS domain are diverse, including numerical, categorical, ordinal, textual, etc. Adding perturbations to these various types of data is not as straightforward as adding perturbations to pixels. Moreover, even features with the same type may have different characteristics; for example, the range of one numerical feature might be completely different from another numerical feature, adding another layer of complexity to crafting the attack.
     \item \textbf{Few Remaining Perturbable Features:} As mentioned in Section \ref{Background}, feature reduction can be a potential defense strategy against adversarial attacks, as it reduces the attack surface and increases the model's robustness. Due to the NIDS feature constraints mentioned in this section, the NIDS domain has fewer perturbable features compared to other domains like computer vision, which act as a natural line of defense against adversarial attacks.
     
\end{itemize}

\subsubsection{Problem-Space Perturbations}
In problem-space adversarial attacks, the attacker modifies the actual packet to deceive the target model. These attacks are much more practical than feature-space attacks, as the attacker only has access to the file and not the feature vector. However, adding perturbations to network packets is extremely difficult.
Usually, a problem-space attack begins with feature-space perturbations. Then, these feature-space manipulations are translated into real-world modifications of network packets, a process referred to as the \textbf{Inverse Feature-Mapping Problem} \cite{pierazzi2020intriguing}. This problem poses a significant barrier for attackers and often complicates, and in most cases, prevents, the direct application of gradient-based feature-space attacks to discover problem-space adversarial examples \cite{pierazzi2020intriguing}. Moreover, there is no guarantee that an optimal solution in the problem space will be similar to the target adversarial features \cite{he2023adversarial}.

Furthermore, the inverse feature process mapping can introduce unintended consequences known as side effect features \cite{pierazzi2020intriguing}. \textbf{Side effect features} refer to the unintended characteristics or features that emerge when applying adversarial perturbations to data. These features arise due to the correlations, limitations, and constraints involved in translating manipulations from the feature space to real-world modifications in the problem space. In simpler terms, side effect features are altered features that arise just to satisfy the constraints of the problem space. It has been observed that these features do not follow any particular direction of the gradient, meaning they can have both positive and negative effects.


On another note, the attacker must know which features and feature transformation techniques are utilized by the target model so that the modification of the packet produces perturbations to the features after the feature extraction phase. This requirement means that \textbf{problem-space attacks inherit all the NIDS feature constraints} discussed in the previous subsection from feature-space attacks, as the attacker must address all of them.

Furthermore, even if the attackers succeed in modifying the packet correctly to trick the model, these modifications will more than likely alter the \textbf{network and malicious functionality} of the perturbed packets \cite{rosenberg2021adversarial}. For example, changing flags or length values in the header will render them inaccurate, and packets with faulty headers are immediately discarded at the destination, resulting in a loss of network functionality. Moreover, adding perturbations to the payload of the packet could change its intended malicious functionality. For instance, if the attackers aim for an SQL injection attack, modifying the packet's content will most likely nullify its hostile purpose. It is worth mentioning that existing problem-space adversarial attacks against ML-NIDS lack evaluation of whether their generated examples maintain malicious functionality \cite{he2023adversarial}. This means current attacks cannot guarantee that the altered packets will still be harmful.

Therefore, as depicted in Figure \ref{fig:Adversarial_Examples}, crafting adversarial attacks against NIDS is much more complicated compared to computer vision. While adding small perturbations to pixels in computer vision may not change their functionality, in the case of NIDS, such attacks can potentially alter both their malicious behavior and network functionality, as shown in Figure \ref{fig:FilePert}.

\subsection{Adaptivity to Dynamic ML}
\label{adaptivity}
 The vast majority of adversarial attacks depend on the gradients of the target models to maximize the loss and make the attacked model produce faulty decisions. On the other hand, as previously mentioned, there is a trend to use dynamic ML which is reasonable to assume that NIDSs would adopt. Because the gradients in dynamic ML are always changing, that could add to the complexity of creating an attack. To test this assumption, we investigated the effect of dynamic learning on adversarial attacks against ML-NIDS. In the next sections, we will discuss our experimental environment and results.

\section{``Practical'' Attacks} \label{Practical}

It is well established that a practical evasion adversarial attack must be black-box, problem-space, while maintaining malicious and network functionality. In addition to these basic prerequisites, based on our literature review, threat modeling (attack tree), and practicality issues taxonomy, we identified three potential nodes or ways to conduct a practical evasion adversarial attack on ML-NIDS: \textbf{(1) Adding random or blind perturbations, (2) side-channel querying, and (3) knowing up-to-date information on training data of the target model to create a surrogate model} This section explores attacks from literature that use these approaches to conduct evasion adversarial attacks on ML-NIDS. We also explore in this section whether they are just practical or whether there are some catches.

\subsection{``Practical'' Attacks from Literature}
\subsubsection{Adding Random or Blind Perturbations}
Apruzzese et al. \cite{apruzzese2024adversarial} present what we consider a practical evaison adversarial attack because it operates under realistic black-box constraints with no model knowledge, uses problem-space perturbations on raw packets, and maintains network/malicious functionality. The attack creates blind perturbations by adding 1-100 random bytes of padding to UDP packets and TCP packets with PSH flags using scapy, ensuring protocol compliance through proper checksum recreation. This approach is practical because it reflects the most constrained attacker scenario: infected hosts within an organization that attackers have compromised via exploits or accidental infection to generate malicious network traffic for objectives like exfiltration or sabotage. These attackers attempt evasion without sophisticated ML-NIDS knowledge while representing physically realizable attacks executable by any attacker controlling network communications.

\begin{table*}
\centering
\caption{Summary of ``Practical'' Evasion Adversarial Attacks Against Network Intrusion Detection Systems}
\label{tab:nids-attacks-comparison}
\adjustbox{width=\textwidth,center}{%
\renewcommand{\arraystretch}{1.1}
\small

\begin{tabular}{@{}p{2.5cm}p{2cm}p{0.6cm}p{2.4cm}p{1.8cm}p{2.2cm}p{2cm}p{2cm}p{2.8cm}@{}}
\toprule
\textbf{Paper \& Authors} & \textbf{Attack Type} & \textbf{Year} & \textbf{Black-box Level} & \textbf{Problem-space} & \textbf{NIDS Output Access} & \textbf{Malicious Function Preserved} & \textbf{Network Function Preserved} & \textbf{Practical Constraints} \\
\midrule

\textbf{Adversarial Perturbations meet Concept Drift \cite{apruzzese2024adversarial}} & 
Blind perturbations under concept drift & 
2024 & 
\textcolor{green}{\textbf{COMPLETE}} - No model knowledge, unaware of drift & 
\textcolor{green}{\textbf{YES}} - Raw packet perturbations & 
\textcolor{green}{\textbf{NONE}} - Blind without feedback & 
\textcolor{green}{\textbf{YES}} - Attack functionality maintained & 
\textcolor{green}{\textbf{YES}} - Protocol compliance preserved & 
Most restrictive: no model knowledge, no drift awareness, real-world deployment \\

\midrule

\textbf{Vulnerability Disclosure through Adaptive Black-Box Attacks on NIDS \cite{ennaji2025vulnerability}   }  & 
Adaptive with side-channel analysis & 
2025 & 
\textcolor{green}{\textbf{COMPLETE}} - No architecture, parameters, or training data access & 
\textcolor{green}{\textbf{YES}} - Metadata manipulation within protocol ranges & 
\textcolor{green}{\textbf{SIDE-CHANNEL}} - CPU, memory, response times only & 
\textcolor{green}{\textbf{YES}} - Attack packet content preserved & 
\textcolor{green}{\textbf{YES}} - Protocol adherence maintained & 
Metadata-only manipulation, Variance Inflation Factor (VIF) filtering preserves correlations, stealth operation \\ 

\midrule

\textbf{Adv-Bot: Realistic Adversarial Botnet Attacks \cite{debicha2023adv}} & 
Substitute model-based botnet attacks & 
2023 & 
\textcolor{green}{\textbf{COMPLETE}} - Very limited target NIDS knowledge & 
\textcolor{green}{\textbf{YES}} - Acts on traffic space respecting domain constraints & 
\textcolor{green}{\textbf{SUBSTITUTE MODEL}} - No queries, uses surrogate models & 
\textcolor{green}{\textbf{YES}} - All intended malicious functionality maintained & 
\textcolor{green}{\textbf{YES}} - Domain constraints respected, valid traffic & 
Surrogate models trained independently, transferability property exploitation, no target interaction \\

\bottomrule
\end{tabular}
}
\end{table*}

\subsubsection{Side-Channel Querying}

Ennaji et al. \cite{ennaji2025vulnerability} present another practical route for evasion adversarial attacks. Their attack operates under strict black-box constraints using only passive side-channel observations like response times, CPU usage, and packet drop rates to infer NIDS behavior. The attack creates realistic perturbations through random walk modifications applied to network metadata (packet length, timestamps, ports) captured via network sniffing tools, ensuring protocol compliance. This approach is practical because it reflects genuine attacker capabilities through silent probing and gradual, natural traffic variations that blend with normal network activity, combined with adaptive feature selection using change-point detection and causality analysis to identify sensitive features without direct system access.

\subsubsection{Knowing up-to-date information on training data of the target model to create a surrogate model}

Debicha et al. \cite{debicha2023adv} also introduced an evasion adversarial attack that operates under realistic black-box constraints with no knowledge of the target NIDS architecture or parameters, using only traffic-based manipulation that directly modifies raw network packets rather than extracted features. The attack achieves practicality through transferability-based evasion, where attackers train surrogate models using sniffed network traffic from their infected botnet machines, then generate adversarial botnet traffic that transfers to the target NIDS without requiring queries or feedback. The approach is realistic because it respects domain constraints through syntactic and semantic validation, manipulating only three directly controllable network factors: duration, packet count, and byte quantities. This is accomplished via time manipulation, packet injection/retention, and byte expansion/compression techniques using tools like Scapy and Hping, while maintaining the underlying malicious functionality of botnet communications.

Table \ref{tab:nids-attacks-comparison} shows a summary of the ``Practical'' Evasion Adversarial Attacks Against Network Intrusion Detection Systems mentioned in this section.

\subsection{The Catch?}

Although we find that these attacks are the most practical, we still think that there is a catch in the practicality of these attacks and some hurdles that attackers must overcome. 

For blind perturbations, they can actually backfire, making adversarial NetFlows easier to detect than their unperturbed malicious variants. Since attackers have no model knowledge and no feedback from the ML-NIDS, there's no guarantee that perturbations will help with evasion and may actually harm their objectives. The attack's effectiveness varies significantly across different ML algorithms and architectures, with some perturbations showing no statistically significant impact on detection rates \cite{elshehaby2026novel}.

Regarding side-channel querying attacks, while practical, there are significant catches limiting effectiveness. The major limitation is dependency on unreliable side-channel indicators that vary based on network load, hardware configurations, and environmental factors beyond attacker control. Additionally, the attack requires extensive computational overhead for change-point detection and causal analysis, making it time-consuming. Moreover, as they require sustained network access over extended periods, this increases detection likelihood by network administrators or advanced monitoring systems.

\begin{figure}
        \centering
        \includegraphics[width=0.4\linewidth,keepaspectratio=true]{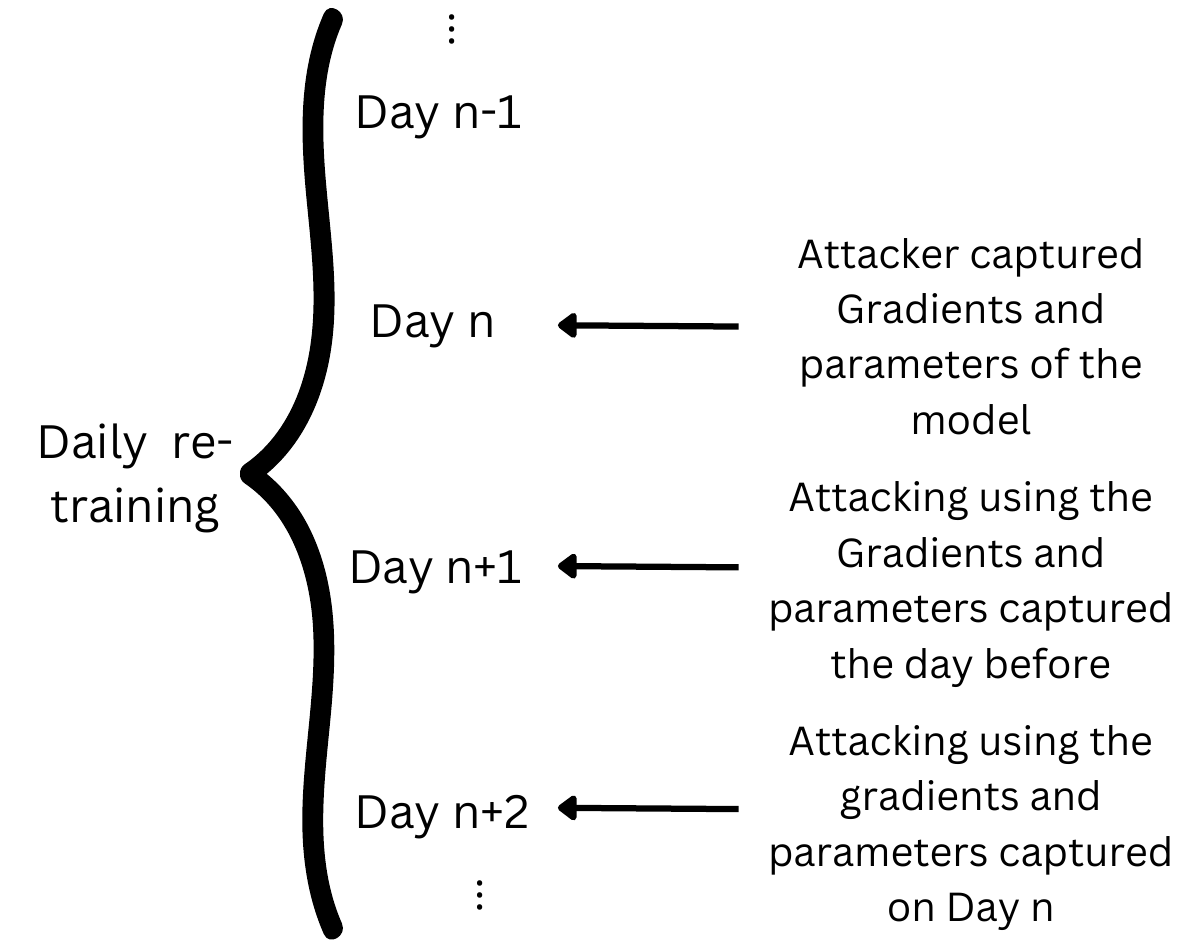}
        \caption{Attacking Scenario with Continuous Training: the impact of adversarial attacks before (attacking in Day n) and after re-training (attacking in Day n+1 and Day n+2)}
        \label{fig:Scenario}
    \end{figure}

While practical, Debicha et al.’s \cite{debicha2023adv} attack faces several significant catches that severely limit its real-world effectiveness. The most critical limitation is the impractical data requirement; the paper assumes attackers have access to 50\% of the target's training dataset, which contradicts realistic black-box scenarios where attackers would never obtain such extensive labeled network traffic data through simple ``network sniffing''. Additionally, transferability uncertainty means there's no guarantee that adversarial instances crafted for surrogate models will successfully evade the actual target NIDS. Domain constraint limitations restrict the perturbation space by requiring maintenance of semantic relationships between network features, while the small perturbation requirement means attacks may fail against robust or ensemble-based NIDS that can detect subtle traffic anomalies. 

Despite the limitations mentioned in this section, we still believe that these attacks could potentially be lethal against NIDS, but we also think that for each of those attacks in their current form, there is still a catch (or catches). That's why we put the word practical in quotation marks.

\section{Investigating of The Effect of Dynamic Learning on Adversarial Attacks Against ML-NIDS}

\label{Investigating_Dynamic}


As noted in Section~\ref{adaptivity}, we believe that adopting a dynamic model might change the effectiveness of adversarial attacks. Thus, we conduct experiments to test the effect of continuous training on the effectiveness of adversarial attacks against ML-NIDS.

\subsection{Attack Scenario}
In this section, we will present the results of our experiments that test the effect of the continuous training aspect of the dynamic learning models on adversarial attacks. Our testing scenario, as depicted in Figure \ref{fig:Scenario}, assumes that attackers have somehow acquired the knowledge needed to calculate gradients and craft their adversarial attacks, despite the challenges discussed in Section \ref{taxo} regarding obtaining information about the target.

Due to the complexity of formulating adversarial attacks against NIDS, there is a time delay (at least one or two days in our experiments) before the attackers initiate the evasion. Meanwhile, the target dynamic ML-NIDS undergoes re-training, rendering the knowledge obtained by the attackers outdated. In our experiments, we aim to test the efficiency of adversarial attacks against ML-based NIDS using up-to-date knowledge of the model (Day n) for evasion versus outdated knowledge after one day of re-training the NIDS (Day n+1) and two days (Day n+2). In other words, we attack the continuously daily trained model using the gradients captured on Day n to attack the model on Day n (up-to-date gradients), Day n+1 (one-day-old gradients), and Day n+2 (two-day-old gradients).

\subsection{The Target ML-based NIDS}
\label{Target_ML-based_NIDS}
\begin{figure}[]
        \centering
        \includegraphics[width=0.7\linewidth,keepaspectratio=true]{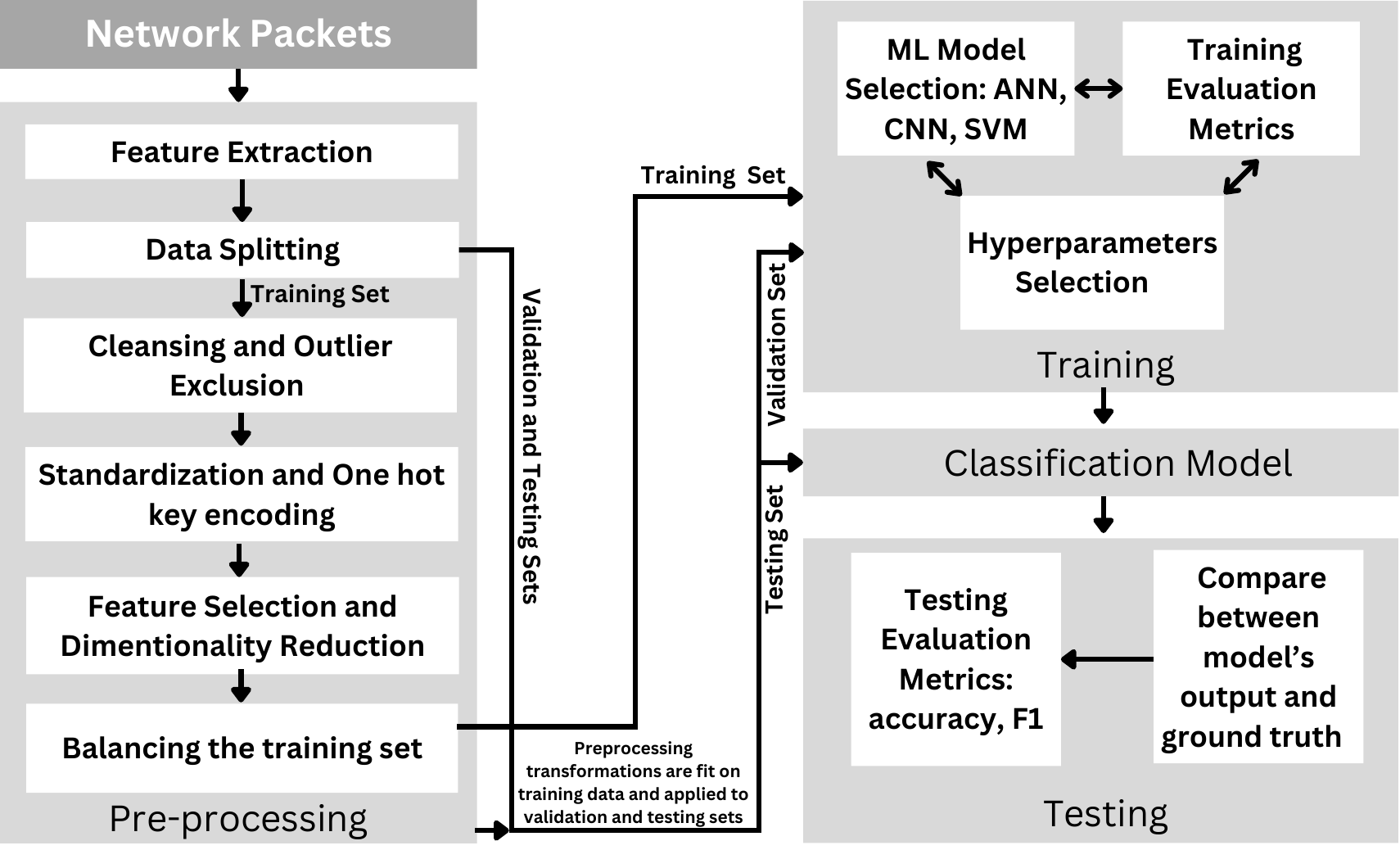}
        \caption{Target ML-based NIDS}
        \label{fig:ML-Based_NIDS}
    \end{figure}

As seen in Figure \ref{fig:ML-Based_NIDS}, the target ML-based NIDS consists of multiple phases. We crafted our own ML-NIDS for our experiments. We will explore our ML-NIDS in the next subsections. 


\subsubsection{Dataset}
The CSE-CIC-IDS2018 \cite{sharafaldin2018toward} dataset was used in our experiments. It was found to be the most up-to-date, and ML models achieve better accuracy when trained and tested on it \cite{ahmad2021network}. Moreover, the CSE-CIC-IDS2018 dataset has a low number of duplicate and uncertain data \cite{karatas2020increasing}. The dataset was created by the Communications Security Establishment (CSE) and the Canadian Institute for Cybersecurity (CIC) in 2018. In the CSV format files, each tuple has 80 features. During the dataset's creation, numerous scenarios were constructed, and data were collected and edited daily. The dataset contains six different classes (labels): Benign, Bot, Brute Force, DoS, Infiltration, and SQL Injection. However, the authors of \cite{liu2022error} found that CSE-CIC-IDS-2018 suffers from numerous issues regarding feature generation and labeling. We agree with their findings, as we discovered a considerable number of outlier samples in the dataset that require cleansing before the standardization and training processes for constructing more robust ML models.

\subsubsection{Pre-Processing}

The pre-processing steps employed in our experiments
contained several key procedures.

Initially, a process of data cleansing was implemented, whereby features containing nulls, infinite values, NaN (Not a Number), or single values were removed. Additionally, thousands of numbers in the dataset were stored in string formats, which had to be converted back into numeric values.

\begin{figure}
\centering
\captionsetup[subfloat]{labelfont=scriptsize,textfont=scriptsize}
\subfloat[{\scriptsize The Frequency of ``Bwd Pkt Len Max'' Feature  Values Before Data Cleansing}]{%
  
\centering
\includegraphics[width=0.5\columnwidth ,keepaspectratio=true]{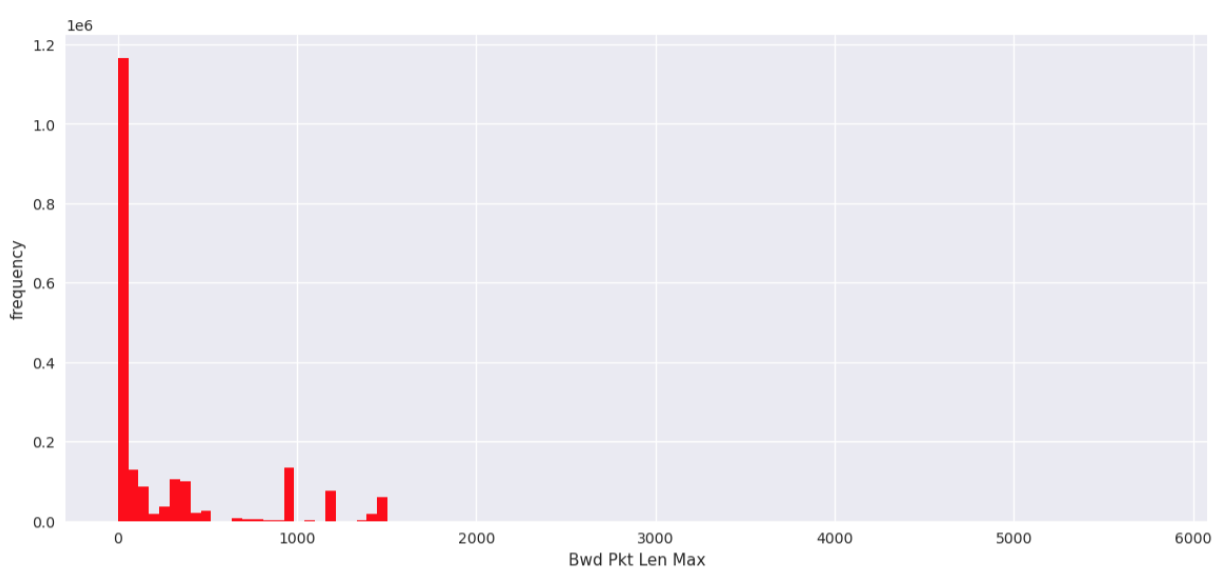}
\centering

    \label{fig:beforeClean}
}

\subfloat[{\scriptsize The Frequency of ``Bwd Pkt Len Max'' Feature  Values After Data Cleansing}]{%

  \centering
\includegraphics[width=0.5\linewidth,keepaspectratio=true]{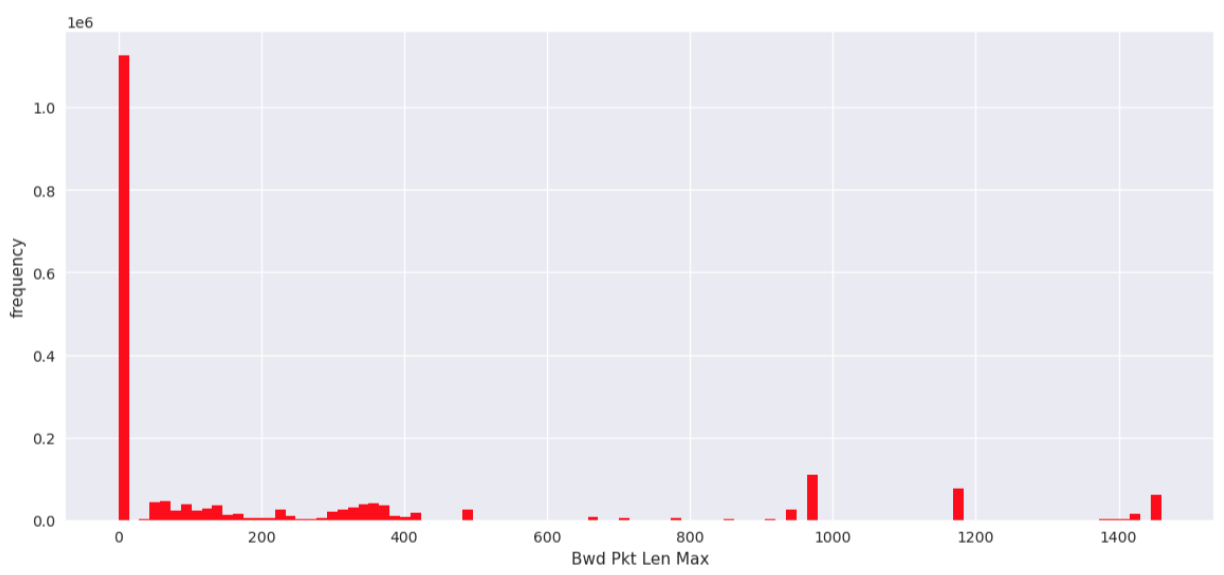}

  \label{fig:AfterCleansing}
}
\caption{The Frequency of ``Bwd Pkt Len Max'' Feature  Values Before and After Data Cleansing}
 \label{fig:BeforeandAfterclean}
\end{figure}

Moreover, we found numerous outlier samples, which are values that deviate significantly from the rest of the dataset. These atypical data points can present challenges for statistical analyses, like mean and standard deviation calculations, which are essential for standardization. Furthermore, in ML training datasets, outliers can disrupt model performance, mislead the training process, lead to non-generalizable models, and increase variance, potentially causing bias and reduced accuracy \cite{ferrari2020dealing}. For example, as seen in Figure \ref{fig:beforeClean}, a minuscule number of records (4 samples out of millions) of ``Bwd Pkt Len Max'' Feature are in the neighborhood of 6000, which is extremely far from the rest of the values. Therefore, we created a method to exclude these outlier samples from the dataset. This technique detects and removes samples with a probability of less than 0.0005 in the cumulative distribution function and those with a probability of more than 0.9995, but only if the number of these records is less than a threshold. The Frequency of ``Bwd Pkt Len Max'' Feature  values Before and After Data Cleansing can be seen in Figure \ref{fig:BeforeandAfterclean}. After excluding the small number of outliers, the range of values for the feature changed from 0 to 6000 to 0 to 1500, as seen in Figure \ref{fig:AfterCleansing}. This outlier-excluding method is applied to all numerical features. Afterward, We performed standardization and one-hot key encoding on the dataset to prepare it for training machine learning models.

\begin{figure}[]
    \centering
    \includegraphics[width=0.44\linewidth,keepaspectratio=true]{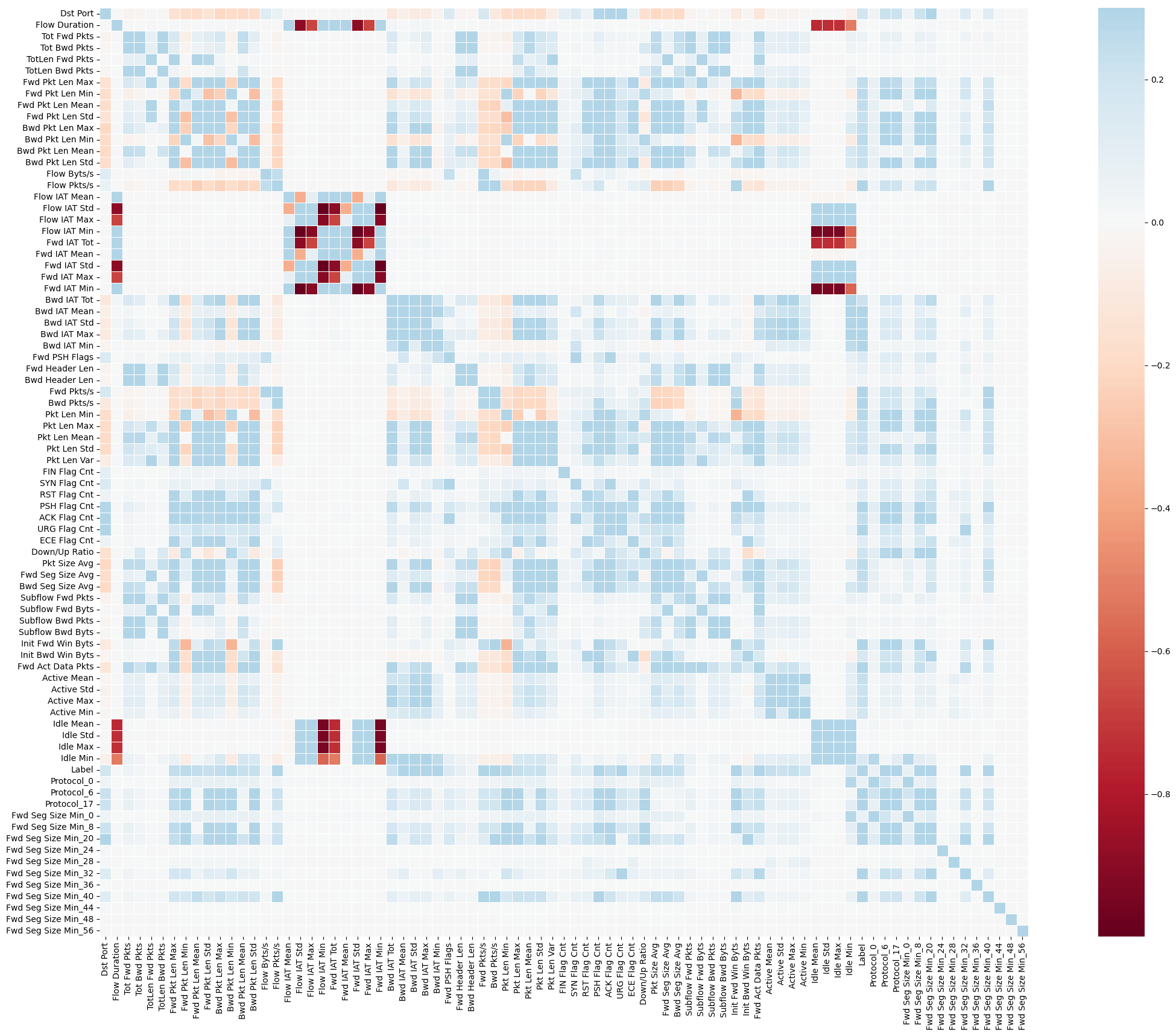}
    \caption{Dataset's Features Correlation Matrix}
    \label{fig:corr}
\end{figure}

Subsequently, as seen in Figure \ref{fig:corr},we conducted a correlation analysis to identify highly correlated features, which were then partitioned into mutually exclusive sets to exclude dependent features. Additionally, we calculated the correlation between features and labels and determined the importance order of features using Random Forests on separate decision trees. Finally, based on the obtained results, we selected the most important features for further analysis.

\begin{figure}[]
\centering
\captionsetup[subfloat]{labelfont=scriptsize,textfont=scriptsize}
\centering
\subfloat[Training Dataset’s Labels Distribution before Random Under-sampling]{%
\includegraphics[width=0.39\linewidth, keepaspectratio=true]{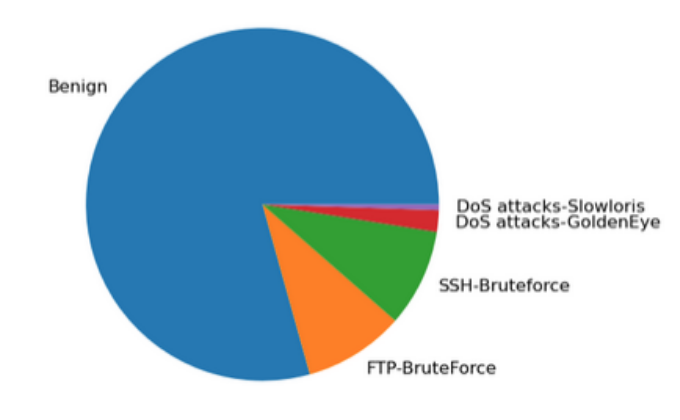}
\label{fig:imb}
}
\subfloat[Training Dataset’s Labels Distribution after Random Under-sampling]{
\includegraphics[width=0.39\linewidth, keepaspectratio=true]{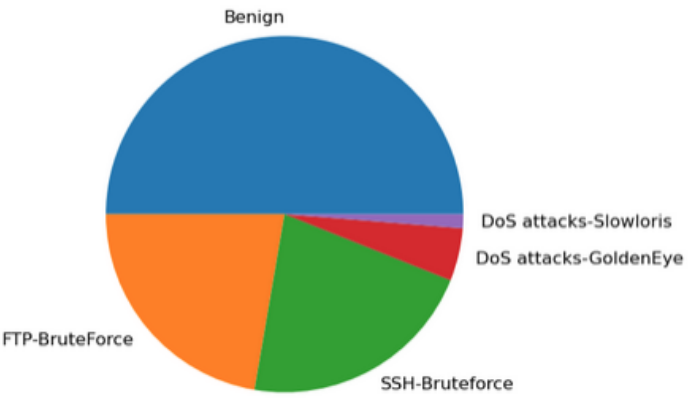}
\label{fig:balanced}
}
\caption{Before and After Balancing the Training Dataset}
\label{fig:Balanced_imbalanced}
\end{figure}

In machine learning, imbalanced datasets can lead to poor accuracy, especially for the minority class \cite{kaur2019systematic}. As seen in Figure \ref{fig:imb}, the used part of the CSE-CIC-IDS2018 dataset was mildly imbalanced, so we balanced it by using the random under-sampling method \cite{liu2020dealing}. The training dataset’s labels distribution after random under-sampling is shown in Figure \ref{fig:balanced}.

\subsubsection{Models' Architectures}

Several ML models were used to create different versions of our experimental NIDS for better comparisons to explore the effect of continuous learning on adversarial attacks.

The first ML model is a fully connected Artificial Neural Network (ANN). The network consists of an input layer containing 50 nodes representing the selected features. Then, two hidden layers with 64 nodes each (using Relu activation function with L2 regularization for the first hidden layer and TANH activation function with L2 regularization for the second hidden layer), which represent the significant attributes describing the nature of the packet, are connected to an output layer with a single node (representing malicious or benign). The ANN uses the ADAM optimizer with an initial learning rate of 0.001. Moreover, the model includes a dropout of 0.4, batch size of 1000, and 20 epochs for training.

Our second model is a Convolution Neural Network (CNN) consisting of three groups of hidden layers. The first group contains a 1D convolution layer with 128 filters, each with a size of 6 and a ReLU activation function. After each convolution layer, we added a dropout layer. The second and third groups are similar to the first, but each contains only 64 filters. Then, we added a max-pooling layer with a pool size of 3 and a stride of 2. After these three groups, we included a flattened layer, two ReLU dense layers (each with 64 nodes), and finally, a Softmax layer.
Our third and last model is a Support Vector Machine (SVM) NIDS. We used the vanilla version with a linear kernel. For all models, we used 80 percent of the selected records from the dataset for training and validation, and the remaining 20 percent for testing.

\subsubsection{Continuous Training }


The CSE-CIC-IDS2018 dataset is partitioned by days, and we assumed that the target ML-NIDS model was re-trained daily (periodic re-training). However, in continuous learning environments, there is a risk of \textbf{catastrophic forgetting}, a phenomenon where a learning model forgets previously learned information when trained on new data or tasks \cite{kirkpatrick2017overcoming}.

To mitigate this issue, we employed a rehearsal-based approach (also known as experience replay) by merging all data frames from previous days for each new day, shuffling the union set, and selecting a random sample with a uniform distribution. This technique reduces catastrophic forgetting by allowing the model to reinforce previously learned patterns while adapting to new data. We repeated this process for three consecutive days to test the impact of continuous training on adversarial attacks.


For our three-day experimental window, this strategy provided sufficient stability (retaining old knowledge) and plasticity (learning new tasks) without requiring more complex or expensive continual learning techniques, serving as a simple yet effective baseline for our experiments.

\subsection{Adversarial Attacks in Our Experiments}

The Adversarial Robustness Toolbox (ART) library \cite{nicolae2018adversarial} was used to generate FGSM \cite{goodfellow2014explaining}, Projected Gradient Descent (PGD) \cite{madry2017towards}, and Basic Iterative Method (BIM) \cite{kurakin2018adversarial} adversarial attacks. We evaluated these attacks by adding adversarial perturbations to the entire test set.

\subsection{Threat Model Assumptions Justification}
\label{sec:Justification}

As previously mentioned, we selected FGSM, PGD, and BIM attacks and periodic re-training for our experiments to evaluate the effect of dynamic learning on adversarial evasion attacks. Our experimental design intentionally tests attacks with the strongest attacker capabilities (white-box access with feature-space manipulation) against the most basic form of dynamic learning (periodic re-training). This combination provides a clear baseline evaluation framework: if the most trivial continual learning approach can effectively reduce the impact of adversarial attacks operating under maximum attacker capabilities, then it establishes strong evidence for dynamic learning's potential effectiveness against more realistic, constrained attack scenarios. Furthermore, Carlini et al. \cite{carlini2017adversarial} specifically recommended using strong iterative attacks (PGD and BIM) in addition to FGSM when evaluating adversarial defense mechanisms, as these attacks provide comprehensive coverage of evasion techniques.

\subsection{Results}

\begin{figure*}
\centering
\captionsetup[subfloat]{labelfont=scriptsize,textfont=scriptsize}
\subfloat[FGSM]{%

\resizebox{0.22\linewidth}{0.1833333\linewidth}{%
\begin{tikzpicture}

\begin{axis}[
    legend pos= south east,
    xmin=0, xmax=40,
    ymin=0, ymax=1.07,
    xtick={5,15,25,35},
    xticklabels={Pre-attack,Day n,Day n+1,Day n+2},   
    ytick={0,0.1,...,1}
            ]
\addplot[mark=*,blue] plot coordinates {
    (5,0.9975)
    (15,0.756)
    (25,0.938)
    (35,0.966)
};
\addlegendentry{ANN}

\addplot[mark=+] plot coordinates {
    (5,0.9975)
    (15,0.8414)
    (25,0.9568)
    (35,0.989)
};
\addlegendentry{CNN}

\addplot[color=red,mark=x]
    plot coordinates {
        (5,0.9975)
        (15,0.1502)
        (25,0.9978)
        (35,0.9978)

    };
\addlegendentry{SVM}
\end{axis}

    \end{tikzpicture}}
\label{fig:FGSM_Graph_ACC}
} \hspace{7mm}
\centering
\subfloat[PGD]{

\resizebox{0.22\linewidth}{0.1833333\linewidth}{%
\begin{tikzpicture}

\begin{axis}[
    legend pos= south east,
    xmin=0, xmax=40,
    ymin=0, ymax=1.07,
    xtick={5,15,25,35},
    xticklabels={Pre-attack,Day n,Day n+1,Day n+2},   
    ytick={0,0.1,...,1}
            ]
\addplot[mark=*,blue] plot coordinates {
    (5,0.9975)
    (15,0.750)
    (25,0.946)
    (35,0.929)
};
\addlegendentry{ANN}

\addplot[mark=+] plot coordinates {
    (5,0.9975)
    (15,0.584)
    (25,0.930)
    (35,0.971)
};
\addlegendentry{CNN}

\addplot[color=red,mark=x]
    plot coordinates {
        (5,0.9975)
        (15,0.150)
        (25,0.998)
        (35,0.998)

    };
\addlegendentry{SVM}
\end{axis}

    \end{tikzpicture}}
\label{fig:ProjectedGradientDescent_Graph_Acc}
}\hspace{7mm}
\centering
\subfloat[BIM]{

\resizebox{0.22\linewidth}{0.1833333\linewidth}{%
\begin{tikzpicture}

\begin{axis}[
    legend pos= south east,
    xmin=0, xmax=40,
    ymin=0, ymax=1.07,
    xtick={5,15,25,35},
    xticklabels={Pre-attack,Day n,Day n+1,Day n+2},   
    ytick={0,0.1,...,1}
            ]
\addplot[mark=*,blue] plot coordinates {
    (5,0.9975)
    (15,0.750)
    (25,0.946)
    (35,0.930)
};
\addlegendentry{ANN}

\addplot[mark=+] plot coordinates {
    (5,0.9975)
    (15,0.688)
    (25,0.932)
    (35,0.979)
};
\addlegendentry{CNN}

\addplot[color=red,mark=x]
    plot coordinates {
        (5,0.9975)
        (15,0.1502)
        (25,0.9978)
        (35,0.9978)

    };
\addlegendentry{SVM}
\end{axis}

    \end{tikzpicture}}
\label{fig:BIM_Graph_Acc}
}
\caption{Accuracy (Y-axis) of the NIDSs before and after the attacks, where Day n represents attacking before re-training, Day n+1 represents attacking one Day after re-training, and Day n+2 represents attacking two days after re-training.}
\label{fig:acc_}
\end{figure*}

\begin{figure*}
\captionsetup[subfloat]{labelfont=scriptsize,textfont=scriptsize}
\centering
\subfloat[FGSM]{%
\resizebox{0.22\linewidth}{0.1833333\linewidth}{%
\begin{tikzpicture}

\begin{axis}[
    legend pos= south east,
    xmin=0, xmax=40,
    ymin=0, ymax=1.07,
    xtick={5,15,25,35},
    xticklabels={Pre-attack,Day n,Day n+1,Day n+2},   
    ytick={0,0.1,...,1}
            ]

\addplot[mark=*,blue] plot coordinates {
    (5,0.9975)
    (15,0.677)
    (25,0.934)
    (35,0.966)
};
\addlegendentry{ANN}

\addplot[mark=+] plot coordinates {
  (5,0.9975)
  (15,0.8627)
    (25,0.9586)
    (35,0.9892)
};
\addlegendentry{CNN}

\addplot[color=red,mark=x]
    plot coordinates {
        (5,0.9975)
        (15,0.0)
        (25,0.9978)
        (35,0.9978)

    };
\addlegendentry{SVM}
\end{axis}

    \end{tikzpicture}}
\label{fig:FGSM_Graph_F1}
} \hspace{7mm}
\centering
\subfloat[PGD]{

\resizebox{0.22\linewidth}{0.1833333\linewidth}{%
\begin{tikzpicture}

\begin{axis}[
    legend pos= south east,
    xmin=0, xmax=40,
    ymin=0, ymax=1.07,
    xtick={5,15,25,35},
    xticklabels={Pre-attack,Day n,Day n+1,Day n+2},   
    ytick={0,0.1,...,1}
            ]
\addplot[mark=*,blue] plot coordinates {
     (5,0.9975)
     (15,0.666)
    (25,0.943)
    (35,0.924)
};
\addlegendentry{ANN}

\addplot[mark=+] plot coordinates {
    (5,0.9975)
    (15,0.689)
    (25,0.933)
    (35,0.971)
};
\addlegendentry{CNN}

\addplot[color=red,mark=x]
    plot coordinates {
        (5,0.9975)
        (15,0)
        (25,0.998)
        (35,0.998)

    };
\addlegendentry{SVM}
\end{axis}

    \end{tikzpicture}}
\label{fig:ProjectedGradientDescent_Graph_F1}
} \hspace{7mm}
\centering
\subfloat[BIM]{
\resizebox{0.22\linewidth}{0.1833333\linewidth}{%
\begin{tikzpicture}

\begin{axis}[
    legend pos= south east,
    xmin=0, xmax=40,
    ymin=0, ymax=1.07,
    xtick={5,15,25,35},
    xticklabels={Pre-attack,Day n,Day n+1,Day n+2},   
    ytick={0,0.1,...,1}
            ]
\addplot[mark=*,blue] plot coordinates {
    (5,0.9975)
    (15,0.667)
    (25,0.943)
    (35,0.925)
};
\addlegendentry{ANN}

\addplot[mark=+] plot coordinates {
    (5,0.9975)
    (15,0.751)
    (25,0.934)
    (35,0.979)
};
\addlegendentry{CNN}

\addplot[color=red,mark=x]
    plot coordinates {
        (5,0.9975)
        (15,0.0)
        (25,0.9978)
        (35,0.9978)

    };
\addlegendentry{SVM}
\end{axis}

    \end{tikzpicture}}
\label{fig:BIM_Graph_f1}
}
\caption{F1-measure (Y-axis) of the NIDSs before and after the attacks, where Day n represents attacking before re-training, Day n+1 represents attacking one day after re-training, and Day n+2 represents attacking two days after re-training.}
\label{fig:F1_}
\end{figure*}

Figures \ref{fig:acc_} and \ref{fig:F1_} demonstrate that continuous re-training might reduce the effect of FGSM, PGD, and BIM attacks across three NIDS architectures (ANN, CNN, SVM). When attacks are launched on Day n using current model knowledge, all classifiers experience significant performance degradation, with SVM suffering the most severe drop. However, at Day n+1, which corresponds to an attack one day after re-training (with knowledge that's one day old), all models recover to over 90\% accuracy and F1-score. By Day n+2, which corresponds to an attack two days after re-training (with knowledge that's two days old), performance remains stable. \textbf{These results suggest that attacks crafted with slightly outdated knowledge of the model might have a lower effect on ML models.} That indicates that continuously training the model might introduce an additional hurdle for attackers, as they would constantly need to obtain the updated gradients of the model, which can be a complex task, especially in the NIDS domain. 
Moreover, it is worth mentioning that this recovery of the model's measures occurred after just one and two re-training sessions. Furthermore, \textbf{these results were achieved without adversarial training}, meaning that the re-training data didn't contain adversarial samples \cite{abou2020evaluation}. For more detailed results, see Appendix \ref{Detailed_Results}.



\subsection{Interpretation of the Results}

\begin{figure}[]
    \centering
    \includegraphics[width=0.871\linewidth,keepaspectratio=true]{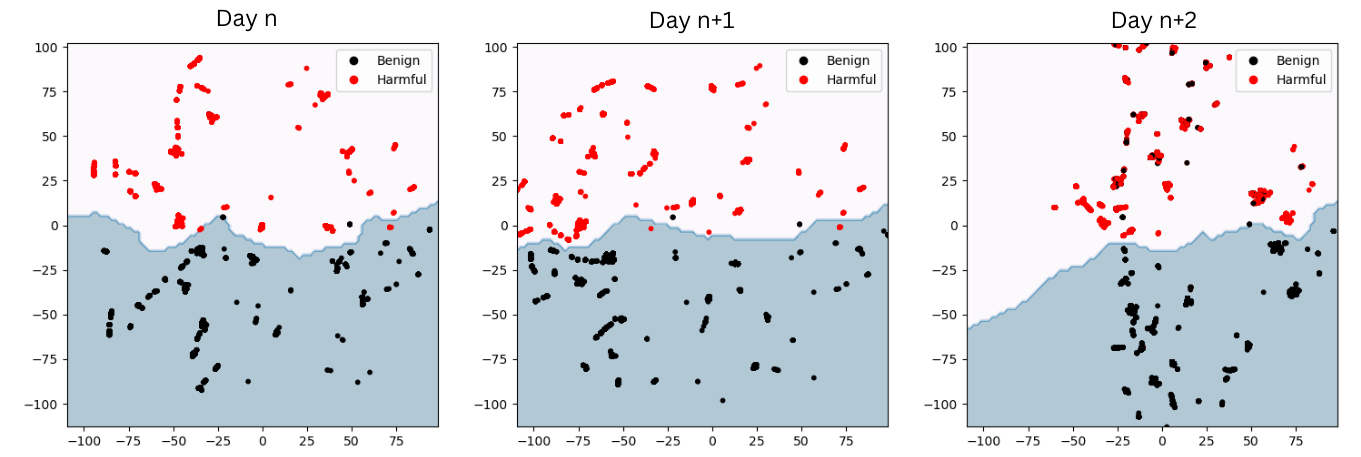}
    \caption{Decision Boundary Evolution using t-SNE}
    \label{fig:tsne}
\end{figure}
\begin{figure}[]
    \centering
    \includegraphics[width=0.871\linewidth,keepaspectratio=true]{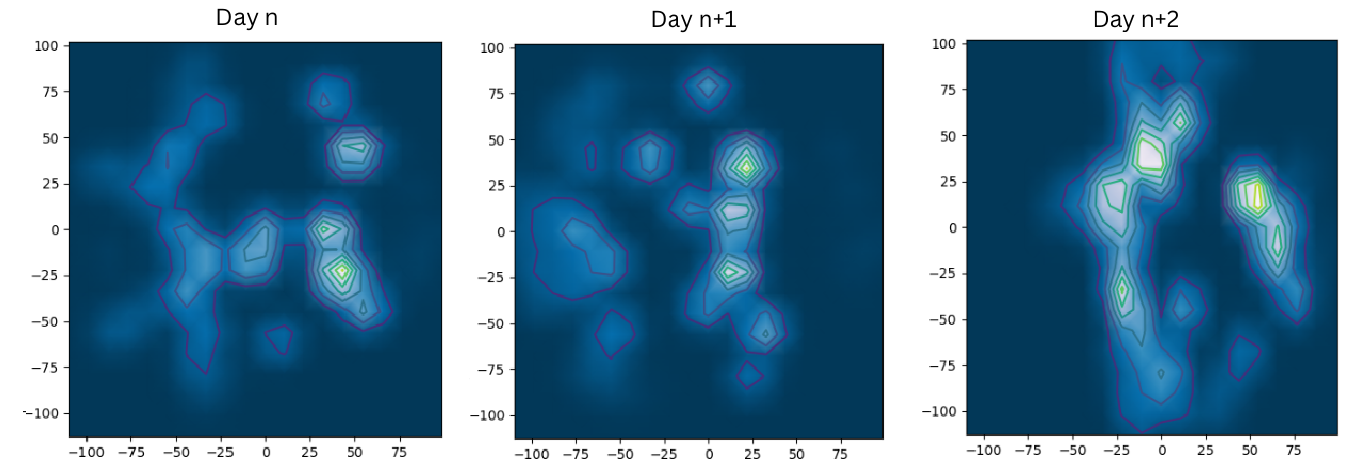}
    \caption{Data Distribution Evolution using t-SNE, wher color intensity encodes data density, with lighter areas representing higher concentrations of points.}
    \label{fig:dens}
\end{figure}

To visualize and gain an intuitive understanding of the data and results, we combine the visualization of Voronoi-based representations of multidimensional decision boundaries in scatterplots with the visualization of distances from data elements to the multidimensional boundary \cite{migut2015visualizing}. Data elements are reduced from 50 dimensions to 2 dimensions using the t-SNE (t-distributed Stochastic Neighbor Embedding) algorithm \cite{van2008visualizing}. In other words, we show the evolution of the optimal multidimensional decision boundaries in 2D after every re-training session, as seen in Figure \ref{fig:tsne}. Moreover, we also visualize the evolution of the data distribution after every re-training, as shown in Figure \ref{fig:dens}. As expected, the decision boundaries and training data distribution are always changing and evolving, as seen in Figures \ref{fig:tsne} and \ref{fig:dens}.


\begin{figure}[]
    \centering
    \includegraphics[width=0.66\linewidth,keepaspectratio=true]{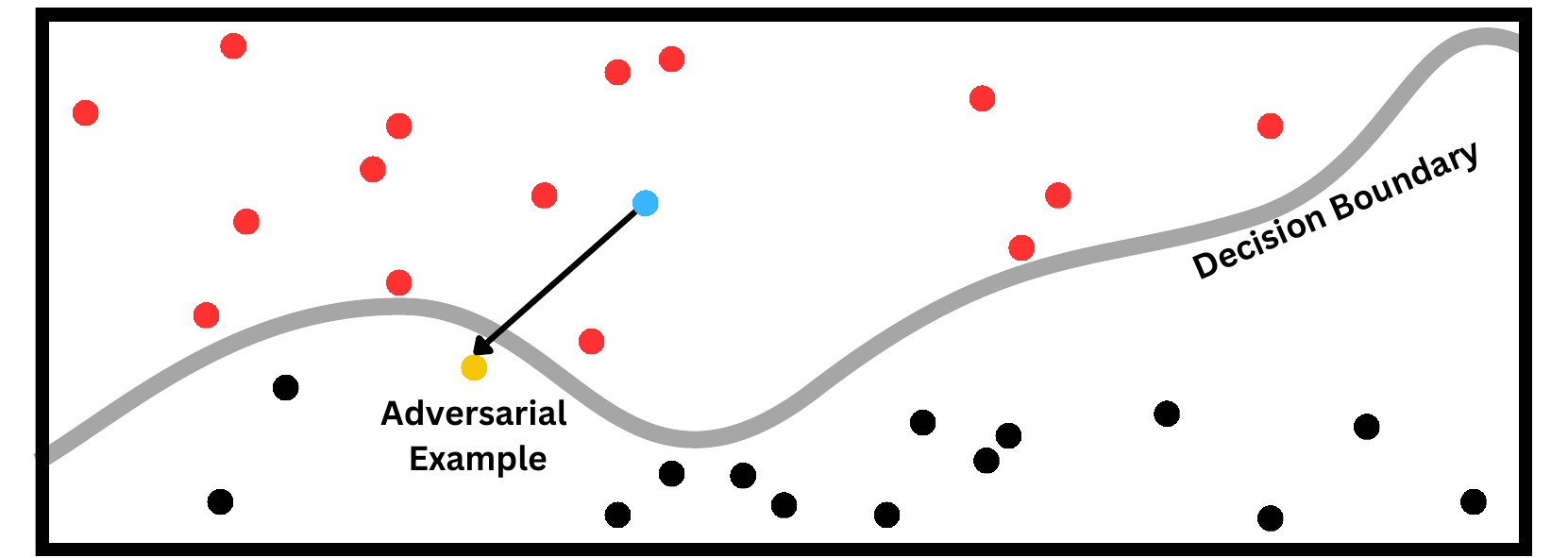}
    \caption{Adversarial Attacks Visualization}
    \label{fig:DB}
\end{figure}

On the other hand, to generate an adversarial example, small perturbations are added to alter the class of a sample. In other words, it attempts to move the sample beyond a nearby decision boundary \cite{heo2019knowledge}. As illustrated in Figure \ref{fig:DB}, perturbations (arrow) are added to a sample (blue circle) to push it into the other class, creating an adversarial example (yellow circle). It has been observed that adversarial samples are typically located near decision boundaries \cite{cao2017mitigating} due to the similarity constraint in adversarial attacks \cite{he2023adversarial}. This constraint aims to ensure that the generated adversarial example remains similar to the original input to a certain degree while still fooling the target model. In simple terms, it limits how much the adversarial example can deviate from the original input while still being effective in fooling the model.

Because adversarial perturbations are highly aligned with a model's weight vectors \cite{waseda2023closer}, when models undergo re-training, these weight vectors shift, rendering previously crafted perturbations misaligned with new decision boundaries. In real-world deployments where models update continuously, attackers face a timing problem: the model they analyze or query (maybe using side-channel querying, which takes time) during attack preparation may differ from the model they encounter at deployment time, a practical constraint rarely considered in research. However, the effectiveness of dynamic learning as a defense depends on multiple factors: model architecture, attack methodology, attacker knowledge, dataset characteristics, and the specific continual learning approach. While no flips, where adversarial samples remain misclassified after updates, or even negative flips may occur, where previously robust inputs become vulnerable after updates \cite{angioni2025robustness} (although Angioni et al. \cite{angioni2025robustness} tested negative flips using current attack knowledge, they could theoretically also occur with outdated adversarial knowledge) this complexity raises a fundamental question: \textbf{if the system is dynamic and decision boundaries continuously shift, crafting adversarial perturbations becomes comparable to adding random noise. This challenges the practicality of employing extremely sophisticated attack methodologies when simpler attack vectors may prove more reliable and as effective in real-world scenarios.}

\section{Conclusion and Future Work}
\label{Conclusion}

In this paper, we extend the research discussion on the practicality of adversarial attacks against NIDS by identifying the questionable practicality prerequisites required to carry out evasion adversarial attacks against NIDS  using an attack tree and by presenting a taxonomy of practicality issues related to ML-based NIDS adversarial attacks. Our attack tree and taxonomy highlight several factors that could make numerous researched adversarial attacks impractical against real-world ML-based NIDSs. Additionally, we identified specific leaf nodes in our attack tree that demonstrate some practicality for real-world implementation and conducted a comprehensive review and exploration of these potentially viable attack approaches from the literature. 

\textbf{It is important to note that our work does not claim that adversarial attacks cannot harm ML-based NIDSs;} on the contrary, we identified three potential routes where attackers can perform practical adversarial evasion attacks on NIDS. However, even these routes still have their own practicality limitations, which is why we find that the gap between research and real-world practicality is wide and deserves to be addressed. We also found that practical threat model assumptions are rare in current research papers, as most attack research still assumes generous threat model assumptions for the attacker. We further argue that this impractical attack research leads to overly expensive defense mechanisms that real-world ML-NIDSs might not actually need. Apruzzese et al. \cite{apruzzese2022position} argue that real-world attackers typically do not rely on gradients or conventional adversarial attack techniques; instead, they employ alternative simple yet effective approaches to evade ML models. If this is the case, then \textbf{the current direction of ML security research needs a significant shift to address real-world challenges adequately.} In the NIDS domain, we find that \textbf{the most practical attacks are simple blind perturbations \cite{apruzzese2024adversarial} that require absolutely no knowledge or querying and are very simple to generate.} Nevertheless, there is no guarantee that these attacks will work consistently in practice.


We also explored the impact of adversarial attacks on dynamic ML-based NIDS, finding that continuous re-training, even without adversarial training, may limit the effect of some adversarial evasion attacks. It is important to note that simple blind perturbations will most likely not be affected by continuous re-training since they do not rely on gradients; however, these attacks remain inconsistent in their lethality and effectiveness. In future research, we plan to expand our attack taxonomy and apply it to network security applications beyond NIDS (e.g., spam detection). We will also explore how more sophisticated dynamic and continuous learning techniques will perform against adversarial attacks.


\section{Acknowledgement}

This work was supported by the Natural Sciences and Engineering Research Council of Canada (NSERC) through the NSERC Discovery Grant program.



\appendix

\section{Open Science}


In alignment with open science policies, this research promotes transparency and reproducibility by:

\begin{enumerate}

    \item \textbf{Dataset:} CSE-CIC-IDS2018  is a publicly available dataset.
    \item \textbf{Code Repository and Supplementary Materials:} Detailed documentation, including instructions for reproducing experiments and analyzing results, as well as the code repository, will be made available upon request.

\end{enumerate}


\section{Detailed Results}
\label{Detailed_Results}

Pre-attack, as seen in Table \ref{Table:Total} (a, b, and c), the NIDSs achieved high accuracy, precision, recall, and F1-score, indicating their effectiveness in detecting intrusions. However, after the FGSM attack, the performance dropped significantly for all three models. Among them, the SVM model experienced the highest drop, with its accuracy and F1 decreasing from 0.997 and 0.997 to 0.150 and 0, respectively. On the other hand, the ANN and CNN models had a decrease in accuracy and F1 of 15 to 30 percent. The reduction in accuracy and F1-measure was particularly prominent, indicating that the attacks successfully evaded the NIDSs' detection. Nevertheless, continuous training improved the NIDSs' resistance to FGSM attacks. On Day n+1 and Day n+2, we observed a significant recovery in accuracy, precision, recall, and F1-score. 

Similarly, as shown in Table \ref{Table:Total} (d to i), the results indicate that, like the FGSM attack, the performance of all three models significantly decreased after the PGD and BIM attacks. Among them, the SVM models experienced the most pronounced decline, while the ANN and CNN models exhibited a milder decrease in accuracy and F1. Continuous training also proved to enhance the NIDSs' resistance to PGD and BIM attacks. On Day n+1 and Day n+2, we observed a substantial recovery in accuracy, precision, recall, and F1-score compared to Day n.

\begin{table*}[]

\small
 \caption{Effect of Continuous Training on FGSM, PGD and BIM Attacks against the ANN, SVM and CNN NIDSs, where Day n represents attacking before re-training, Day n+1 represents attacking one day after re-training, and Day n+2 represents attacking two days after re-training.}
\label{Table:Total}
\noindent\hspace*{-2.2cm}\begin{minipage}{\dimexpr\textwidth+2.2cm\relax}
\begin{center}
\scriptsize
\begin{tabular}{ ||c | c | c | c |c |c|c| c | c | c  || }

  \hline\hline

   \hline

 \multicolumn{5}{|c|}{\textbf{(a) Attack: FGSM - Target Model: ANN}} & \multicolumn{5}{|c|}{\textbf{(b) Attack: FGSM - Target Model: SVM}}\\
 \hline
   & \textbf{Accuracy} & \textbf{Precision} & \textbf{Recall} & \textbf{F1} & &\textbf{Accuracy} & \textbf{Precision} & \textbf{Recall} & \textbf{F1} \\ 
  \hline\hline
 \hline
\textbf{ Pre-attack} & 0.997 & 0.995 & 1 &0.997 &\textbf{Pre-attack} & 0.998 & 0.996 & 1 &0.998\\   
 \hline
  \textbf{Day n} & 0.756 & 0.999 & 0.513 & 0.677 &\textbf{Day n}  & 0.150 & 0 & 0 & 0\\ 
  			
 \hline
 			
  \textbf{Day n+1} &0.938  & 0.996 & 0.880 &0.934  &\textbf{Day n+1} & 0.998 & 0.996 & 1 & 0.998 \\
  			
 \hline
  \textbf{Day n+2} &0.966  & 0.996 & 0.896 &0.966 & \textbf{Day n+2} & 0.998 & 0.996 & 1 & 0.998\\ 

   \hline
   
  \hline\hline

   \hline

   \hline

 \end{tabular}
\end{center}

\begin{center}
\scriptsize
\begin{tabular}{ ||c | c | c | c |c |c|c| c | c | c  || }

  \hline\hline

   \hline

 \multicolumn{5}{|c|}{\textbf{(c) Attack: FGSM - Target Model: CNN}} & \multicolumn{5}{|c|}{\textbf{(d) Attack: PGD - Target Model: ANN}}\\
 \hline
   & \textbf{Accuracy} & \textbf{Precision} & \textbf{Recall} & \textbf{F1} & &\textbf{Accuracy} & \textbf{Precision} & \textbf{Recall} & \textbf{F1} \\ 
  \hline\hline
 \hline
 \textbf{Pre-attack} &  0.997 & 0.994 & 1 &0.997 &\textbf{Pre-attack}  &  0.997 & 0.995 & 1 &0.997\\   
 \hline
  \textbf{Day n} & 0.842 & 0.762 & 0.995 & 0.863 &\textbf{Day n} & 0.750 & 0.999 & 0.501 & 0.666  \\ 
  			
 \hline
 			
  \textbf{Day n+1} &0.957  & 0.921 & 0.999 & 0.959 &\textbf{Day n+1} & 0.946  & 0.996 & 0.896 & 0.943 \\
  			
 \hline
  \textbf{Day n+2} &0.989  & 0.981 & 0.997 & 0.989  &\textbf{ Day n+2 }& 0.929  & 0.995 & 0.861 &0.924 \\ 

   \hline
   
  \hline\hline

   \hline

 \end{tabular}
\end{center} 

\begin{center}
\scriptsize
\begin{tabular}{ ||c | c | c | c |c |c|c| c | c | c  || }

 \hline\hline

   \hline

 \multicolumn{5}{|c|}{\textbf{(e) Attack: PGD - Target Model: SVM}} & \multicolumn{5}{|c|}{\textbf{(f) Attack: PGD - Target Model: CNN}}\\
 \hline
   & \textbf{Accuracy} & \textbf{Precision} & \textbf{Recall} & \textbf{F1} & &\textbf{Accuracy} & \textbf{Precision} & \textbf{Recall} & \textbf{F1} \\ 
  \hline\hline
 \hline
 \textbf{Pre-attack} & 0.998 & 0.996 & 1 &0.998 &\textbf{Pre-attack}  & 0.997 & 0.994 & 1 &0.997\\   
 \hline
  \textbf{Day n} & 0.150 & 0 & 0 & 0  &\textbf{Day n} &  0.584 & 0.551 & 0.920 & 0.689 \\ 
  			
 \hline
 			
 \textbf{ Day n+1} &0.998 & 0.996 & 1 & 0.998 &\textbf{Day n+1} & 0.930  & 0.900 & 0.969 & 0.933\\
  			
 \hline
  \textbf{Day n+2} &0.998 & 0.996 & 1 & 0.998 & \textbf{Day n+2}& 0.971  & 0.961 & 0.982 &0.971\\ 

   \hline
   
  \hline\hline

 \end{tabular}
 
\end{center} 

\begin{center}
\scriptsize
\begin{tabular}{ ||c | c | c | c |c |c|c| c | c | c  || }

  \hline\hline

   \hline

 \multicolumn{5}{|c|}{\textbf{(g) Attack: BIM - Target Model: ANN}} & \multicolumn{5}{|c|}{\textbf{(h) Attack: BIM - Target Model: SVM}}\\
 \hline
   & \textbf{Accuracy} & \textbf{Precision} & \textbf{Recall} & \textbf{F1} & &\textbf{Accuracy} & \textbf{Precision} & \textbf{Recall} & \textbf{F1} \\ 
  \hline\hline
 \hline
 \textbf{Pre-attack} & 0.997 & 0.995 & 1 &0.997 &\textbf{Pre-attack}  & 0.998 & 0.996 & 1 &0.998\\   
 \hline
  \textbf{Day n} & 0.750 & 0.999 & 0.501 &0.667  &\textbf{Day n }& 0.150 & 0 & 0 &0\\ 
  			
 \hline
 			
  \textbf{Day n+1} &0.946 & 0.996 & 0.895 &0.943 &\textbf{Day n+1} & 0.998  & 0.996 & 1 &0.998  \\
  			
 \hline
 \textbf{ Day n+2} &0.930  & 0.995 & 0.863 &0.925 & \textbf{Day n+2} & 0.998  & 0.996 & 1 &0.998\\ 

   \hline
   
  \hline\hline

 \end{tabular}
\end{center} 
\begin{center}
\scriptsize
\begin{tabular}{ ||c | c | c | c |c || }

 \hline\hline

 \hline
 \multicolumn{5}{|c|}{\textbf{(i) Attack: BIM - Target Model: CNN}} \\
 \hline
   & \textbf{Accuracy} & \textbf{Precision} & \textbf{Recall} & \textbf{F1}  \\ 
  \hline\hline
 \hline
\textbf{Pre-attack} & 0.997 & 0.994 & 1 &0.997\\  
 \hline
 \textbf{ Day n} & 0.688 & 0.626 & 0.939 &0.751 \\ 
  			
 \hline
 \textbf{ Day n+1} &0.932  & 0.902 & 0.969 &0.934 \\ 
  			
 \hline
  \textbf{Day n+2} & 0.979  & 0.972 & 0.987 & 0.979 \\ 
  
  \hline\hline

   \hline
\end{tabular}
\end{center}
\end{minipage}
\end{table*}



\end{document}